\documentclass[12pt,letterpaper,english]{article}

 \usepackage{setspace,graphics}
 \usepackage[dvips]{epsfig} 
 \usepackage{a4,amssymb,epsfig,array,cite}
 \usepackage[hypertex]{hyperref}

\usepackage{amsfonts,bm,amssymb,euscript,array,babel}

\newcounter{multieqs}

\newcommand{\be}{\begin{equation}}
\newcommand{\ee}{\end{equation}}
\newcommand{\eq}[1]{(\ref{#1})}

\def\nn{\nonumber}
\def\bea{\begin{eqnarray}}
\def\eea{\end{eqnarray}}
\def\obar{\overline}

\def\beqa{\begin{eqnarray}} 
\def\eeqa{\end{eqnarray}} 
\def\beq{\begin{equation}} 
\def\eeq{\end{equation}}

\def\Tr{{\rm Tr}}

\def\a{\alpha}          
\def\b{\beta}           
  \def\C{\Gamma}  
\def\d{\delta}

\def\g{\gamma}

\def\l{\lambda} \def\L{\Lambda}

  \def\cC{{\cal C}}
  
\def\cG{{\cal G}} \def\cH{{\cal H}} 
  
\def\cM{{\cal M}}  
  \def\cR{{\cal R}}
  \def\cU{{\cal U}}

\def\R{{\mathbb R}}
\def\C{{\mathbb C}}

\def\one{\mbox{1 \kern-.59em {\rm l}}}

\def\msu{\mathfrak{s}\mathfrak{u}}

\def\bit{\begin{itemize}}
\def\eit{\end{itemize}}

\def\({\left(}
\def\){\right)}
\def\diag{\mbox{diag}}

\def\d{\delta}

\def\uno{\mbox{1 \kern-.59em {\rm l}}}

\def\Box{\square}

\def\bcomment#1{}

\renewcommand{\title}[1]{\vspace{10mm}\noindent{\Large{\bf #1}}\vspace{8mm}}
\newcommand{\authors}[1]{\noindent{\large #1}\vspace{5mm}}
\newcommand{\address}[1]{{\itshape #1\vspace{2mm}}}

\begin{document}


\begin{flushright}
UWTHPh-2009-07\\
\end{flushright}

\begin{center}

\title{On the Newtonian limit of emergent NC gravity \\[1ex]
and long-distance corrections \\[1ex]
 }

 \authors{Harold {\sc Steinacker}${}^{1}$}

 \address{Department of Physics, University of Vienna \\
 Boltzmanngasse 5, A-1090 Wien, Austria}

\footnotetext[1]{harold.steinacker@univie.ac.at}

\vskip 1.5cm

\textbf{Abstract}

\vskip 3mm 

\begin{minipage}{14cm}%

We show how Newtonian gravity emerges
on 4-dimensional non-commutative spacetime branes
in Yang-Mills matrix models. 
Large matter clusters such as galaxies are 
embedded in large-scale harmonic deformations of the space-time brane,
which screen gravity for long distances.
On shorter scales, the local matter distribution 
reproduces Newtonian gravity via local 
deformations of the brane and its metric.  
The harmonic ``gravity bag'' 
acts as a halo with effective positive energy density.
This leads in particular to a significant enhancement of  the 
orbital velocities around galaxies at large distances 
compared with the Newtonian case, before dropping to zero 
as the geometry merges with a Milne-like cosmology. 
Besides these ``harmonic'' solutions, there is 
another class of solutions which is more similar to 
Einstein gravity.
Thus the IKKT model provides an accessible candidate for 
a quantum theory of gravity.

\end{minipage}

\end{center}


\setcounter{page}0
\thispagestyle{empty}
\newpage

\begin{spacing}{.3}
{
\noindent\rule\textwidth{.1pt}            
   \tableofcontents
\vspace{.6cm}
\noindent\rule\textwidth{.1pt}
}
\end{spacing}


\section{Introduction}

The aim of this paper is to study the physical properties of 
the effective (``emergent'') gravity which arises in matrix models
such as the IKKT model, with emphasis on 
long-distance and cosmological aspects.

The presently accepted description for cosmology, known as
$\L$CDM model, is able to reproduce the basic observational
data. Its is based on the assumption that gravity is described
by general relativity (GR), leading to the Friedmann equations
which describe a homogeneous and isotropic universe. 
However, this requires a very delicate fine-tuning 
of the basic parameters in order
to reproduce the observational constraints. The most 
problematic aspect is dark energy or
the cosmological constant $\L$, which must be
fine-tuned to 
$\L \approx 2 \times 10^{-3} eV$ on order to produce
the apparent cosmic acceleration inferred from the type Ia supernovae
data. The standard model
also requires significant amounts of dark matter
in order to reconcile the 
galactic rotation curves with GR: The rotation
velocities of stars near or
outside the visible bulk of typical galaxies does not 
decrease with radius as implied by GR 
resp. Newtonian gravity, rather it is more-or-less flat
or even increasing. This is usually explained by postulating
large dark matter halos around the galaxies,
which however has never been observed directly.

These and other problems provide sufficient motivation to look
for alternative descriptions of gravity, in particular 
at large distances. Indeed in spite of its undisputed success,
general relativity has never been
tested directly on cosmological scales. 
Finally, all attempts to define a quantum version of GR 
are faced with serious  conceptual and technical difficulties.

In this paper, we show that matrix models, notably the 
IKKT matrix model \cite{Ishibashi:1996xs}, 
provide a very interesting alternative
theory of gravity which might allow to resolve these problems.
The geometric mechanism for gravity in these Yang-Mills matrix 
models was clarified recently 
\cite{Steinacker:2007dq,Steinacker:2008ri,Steinacker:2008ya},
realizing related ideas 
\cite{Rivelles:2002ez,Yang:2006mn,Yang:2008fb,Muthukumar:2004wj}
in a concise framework. Space-time
is described in terms of a 3+1-dimensional 
noncommutative (NC) brane solution embedded in $\R^{10}$, 
i.e. a quantized space  $\cM_\theta \subset \R^{10}$
which carries a non-degenerate Poisson tensor $\theta^{\mu\nu}(x)$.
All matter and gauge fields live on this 
space-time brane, and there are {\em no} physical fields
propagating 
in the ambient 10-dimensional space\footnote{unlike in
other braneworld scenarios such as 
\cite{ArkaniHamed:1998rs,Dvali:2000hr}}.
An effective dynamical metric $G^{\mu\nu}(x)$ arises on this 
space-time brane, which governs the kinetic term of
all fields more-or-less as in GR. 
This metric is composed of the embedding metric 
$g_{\mu\nu}$ and the Poisson tensor, 
$G^{\mu\nu} \sim \theta^{\mu\mu'}\theta^{\nu\nu'} g_{\mu'\nu'}$,
and defines the effective gravity seen by matter and fields
as in GR. However, the {\em dynamics} of the 
effective metric is not described by the 
Einstein equations.

An essential difference to general 
relativity is that the metric is not a fundamental degree of 
freedom, but arises effectively 
in terms of scalar fields describing the embedding
of space-time $\cM_\theta \subset \R^{10}$, and the Poisson tensor
$\theta^{\mu\nu}$ describing noncommutativity. 
This  makes the dynamics of emergent NC gravity 
somewhat difficult to disentangle.
In this paper, we obtain approximate solutions which 
correspond to static and somewhat localized matter distributions,
having in mind galaxies and their stars inside. 
The bottom line is that large matter clusters such as 
galaxies are embedded in ``gravity bags'' or halos,  
which are deformations of the embedding 
$\cM_\theta \subset \R^{10}$ with very long 
(galactic or cosmological) 
wavelength. These gravity bags turn out to screen
gravity at large  distances, and 
enclose an effective positive ``vacuum energy''. 
Localized matter distributions such as stars then induce
local deformation of this large-scale embedding, which
leads to Newtonian gravity within these gravity bags 
at shorter scales. 

The effective vacuum energy inside the gravity bags
can be considerably larger than the currently 
preferred value in the $\Lambda$CDM model, which -- along with 
other contributions -- leads to a significant enhancement of the 
(galactic) rotation velocities at large distances. This might 
provide an explanation of the observed galactic rotation curves
without invoking large amounts of dark matter. 
Moreover, the effective Newton constant is 
determined by the large-scale deformation resp. 
the gravitational background, and therefore can vary 
somewhat in different locations of the universe. 

The screening of gravity and the different
physics of vacuum energy naturally leads to 
a consistent cosmological picture, where 
the localized matter distributions are embedded in a 
Milne-like cosmology \cite{Klammer:2009ku}.
This is in remarkably good agreement with basic observations, 
requiring considerably less fine-tuning than in the standard model. 
We briefly recall this cosmological solution 
in section \ref{sec:cosmology},  and show how the present
results naturally fit into this cosmological context.

Due to the nonlinear nature of the problem,
the results of this paper are somewhat incomplete and 
preliminary. However, the basic results concerning the
Newtonian limit and the long-distance deviations from 
Newtonian gravity and GR 
are expected to be reliable. 
At short distances the approximations are less 
reliable, and a more complete analysis is required before
seriously addressing e.g. the solar system precision tests.
But in any case, it is remarkable how naturally emergent 
NC gravity seems to reproduce most of the basic observations,
with far less fine-tuning than in the $\Lambda$CDM model.
Given the rigidity of the model and its 
perspective to define a full quantum theory of fundamental
interactions, it certainly deserves to be studied very
thoroughly.

\section{The Matrix Model}
\label{sec:basic}

We consider the following type of Yang-Mills matrix models 
\be
S_{YM} = - \Lambda_{0}^4\, \Tr \( \frac 14 [X^a,X^b] [X^{a'},X^{b'}] 
\eta_{aa'}\eta_{bb'} \,\, + \,\, \frac 12\, \obar\Psi  \Gamma_a
[X^a,\Psi]\) 
\label{YM-action-1}
\ee
where $\eta_{aa'} = \diag(-1,1,...,1)$;
the Euclidean version of the model is obtained 
by replacing $\eta_{aa'}$ with $\delta_{aa'}$.
The degrees of freedom of this model are 
hermitian\footnote{in the Minkowski case, 
we will assume that the time-like matrices are anti-hermitian, see
below. This is consistent with a real metric and action, 
and will be addressed in more detail elsewhere.} 
matrices $X^a\,\in Mat(\infty,\C)$ for $a=0,1,2,..., D-1$, as well as
Grassmann-valued matrices
$\Psi$ which are spinors of $SO(D-1,1)$ resp. $SO(D)$.
The $\Gamma_a$ generate the Clifford algebra in $D$ dimensions.
We introduced also an energy scale $\Lambda_0$ 
which gives the matrices $X^a$ the
dimension of length.
The action is invariant under the fundamental gauge symmetry
\be
X^\mu \to U^{-1} X^\mu U, \quad \Psi \to U^{-1} \Psi U
\qquad U \in \cU(\cH) 
\label{gauge}
\ee
as well as a global $ISO(D-1,1)$ resp. $ISO(D)$
symmetry, where
translations act as $X^a \to X^a + c^a \one$. 
However there is no space-time 
or geometry to start with; space and geometry ``emerge'' only 
as solutions or backgrounds of the model. The models
can be obtained as dimensional reduction of
large-$N$ super-Yang-Mills theory to dimension zero.
The IKKT model with $D=10$ is singled out by an extended 
matrix supersymmetry \cite{Ishibashi:1996xs}. 

It is easy to see how space(time) arises in such a  
model\footnote{This basic observation has been made
by many authors including 
\cite{Banks:1996nn,Chepelev:1997ug,Nair:1998bp,Aoki:1999vr}, and 
it is obvious from the point of view of NC gauge theory 
\cite{Alekseev:2000fd,Douglas:2001ba,Szabo:2001kg}.}.
Dropping the fermionic terms for now, 
the equations of motion 
\be
[ X^a,[X^b,X^{a'}] ]\eta_{aa'} = 0 
\label{eom-vac-matr}
\ee
admit in particular $4$-dimensional noncommutative or quantum spaces 
$\cM_\theta \subset \R^{{10}}$ as solution. 
 This means that we can split the set of matrices as 
\be
X^a = (X^\mu,\phi^i), \qquad \mu = 0,...,3, 
\,\,\, i=4, ..., 9 
\label{extradim-splitting}
\ee
where the 4 generators  $X^\mu$ 
are assumed to generate the full matrix algebra 
$Mat(\infty,\C) $, which is interpreted as
space of (noncommutative)  functions on $\cM$,
i.e. $Mat(\infty,\C) \cong \cC_\theta(\cM)$. This is the basic
idea of noncommutative geometry. 
The ``scalar fields'' $\phi^i=\phi^i(X^\mu)$ are 
then functions of $X^\mu$. The prototype
of such a solution is the Moyal-Weyl quantum plane
where  $[X^\mu,X^\nu] = i \obar\theta^{\mu\nu}\one, \,\, \phi^i = 0$,
but we will focus on the case of nontrivial $\phi^i$ here.

The basic hypothesis in is that space-time is described by such 
a quantum space solution of \eq{eom-vac-matr}. 
We focus on the {\em semi-classical}  
limit of such quantum spaces, indicated by $\sim$.
Then  the
$\phi^i(x)$ define the embedding of a 4-dimensional 
submanifold\footnote{cf. \cite{Chaichian:2006ht} for a
different approach to NC submanifolds.} 
$\cM \subset \R^{10}$, and
\be 
[X^\mu,X^\nu] \sim i\theta^{\mu\nu}(x), \quad  \mu,\nu = 1,..., 4
\label{theta-induced}
\ee 
can be interpreted as Poisson structure on $\cM$.
In particular, the matrices $X^\mu \sim x^\mu$ are 
interpreted as quantization of 
coordinate functions on $\cM$. Thus the matrix model provides
preferred coordinates $x^\mu$, which however have no physical 
meaning whatsoever. From the point of view of 
GR, they essentially ``fix the gauge'',
disposing of diffeomorphism invariance which does not 
make sense in the matrix model.
Since gauge-dependent objects are always unphysical, this 
has no implications on the physical content of the model.

All physical fields in this model arise from
fluctuations in the matrix model
around such a background (leading to nonabelian\footnote{In the
nonabelian case the background solution
is generalized to include a $\msu(n)$ factor, 
see e.g. \cite{Steinacker:2008ya,Madore:2000en}} gauge fields
and scalars)
and from the fermionic matrices $\Psi$. Since 
$Mat(\infty,\C)\cong \cC_\theta(\cM)$  by assumption, 
it follows that they all
live only on the brane $\cM$, and there is no 
physical higher-dimensional ``bulk''
which could carry any propagating degrees of freedom.
This does not exclude the existence of compactified physical
extra dimensions in the matrix model, but these are different 
backgrounds which we will leave aside for simplicity here.

\paragraph{Emergent geometry.}

The Poisson tensor $\theta^{\mu\nu}(x)$ not only
governs the noncommutative structure of $\cM$, it
also plays a crucial but implicit role in the low-energy effective action
and the metric on $\cM$. We need to assume that it is non-degenerate,
so that its inverse 
\be
\theta^{-1}_{\mu\nu}(x)\, 
\ee
defines a symplectic 2-form on $\cM$. Then
the trace on $Mat(\infty,\C)$ is given semi-classically by the 
volume of this symplectic form,
\bea
(2\pi)^{2}\, Tr f &\sim& \int d^{4} x\, \rho(x)\, f \nn\\
\rho(x) &=& (\det\theta^{-1}_{\mu\nu})^{1/2} .
\label{rho-def-general}
\eea
We can now  extract the semi-classical limit
of the matrix model and its physical meaning.
To understand the effective geometry of $\cM^{4}$,
consider a test-particle on $\cM^{4}$, 
modeled by a scalar field $\varphi$ for simplicity
(this could be e.g. an $su(k)$ component of $\phi^i$).
In order to preserve gauge invariance, 
the kinetic term must have the form
\bea
S[\varphi] &\equiv& - Tr [X^a,\varphi][X^b,\varphi] \eta_{ab} = 
- Tr \([X^\mu,\varphi][X^\nu,\varphi] \eta_{\mu\nu} 
  + [\phi^i,\varphi][\phi^j,\varphi] \delta_{ij}\) . 
\label{MM-action-scalar}
\eea
Expressing the $\phi^i$ in terms of $X^\mu$ and using
\be
[\phi^i,f(X^\mu)] \,\sim\, i\theta^{\mu\nu}\partial_\mu \phi^i \partial_\nu f
\ee
this kinetic term can be cast into covariant form
\bea
S[\varphi]
 &\sim&\frac 1{(2\pi)^2}\, \int d^{4} x\; 
|G_{\mu\nu}|^{1/2}\,G^{\mu\nu}(x)
 \partial_{\mu} \varphi \partial_{\nu} \varphi \,,
\label{covariant-action-scalar}
\eea
where \cite{Steinacker:2008ri}
\bea  
G^{\mu\nu}(x) &=& e^{-\sigma}\,\theta^{\mu\mu'}(x) \theta^{\nu\nu'}(x) 
 g_{\mu'\nu'}(x)  
\label{G-def-general}  \\
e^{-\sigma} &=& \rho\, |g_{\mu\nu}(x)|^{-\frac 12} , 
\label{sigma-rho-relation} \nn\\
g_{\mu\nu}(x) &=& \partial_\mu x^a \partial_\mu x^a \,\,
= \eta_{\mu\nu}(x)  + \partial_\mu \phi^i \partial_\mu \phi^j \d_{ij}.
\label{g-explicit}
\eea
Here $g_{\mu\nu}(x)$ 
is the metric induced on $\cM^{4}\subset \R^{10}$ via 
pull-back of $\eta_{ab}$.
Therefore the kinetic term for $\varphi$ on  $\cM^{4}_\theta$
is governed by the metric $G_{\mu\nu}(x)$, 
which depends on the Poisson tensor $\theta^{\mu\nu}$ and the 
embedding 
metric $g_{\mu\nu}(x)$. It turns out that
the same metric also governs nonabelian gauge fields 
\cite{Steinacker:2008ya} and fermions \cite{Steinacker:2008ri} 
in the matrix model, hence
$G_{\mu\nu}$ must be interpreted as
gravitational metric. There is no need and no room 
for invoking any ``principles''.

We note that 
\be
|G_{\mu\nu}(x)| = |g_{\mu\nu}(x)| ,
\label{G-g-4D}
\ee
which means that the Poisson tensor $\theta^{\mu\nu}$ does 
not enter the Riemannian volume at all. This is 
important for stabilizing flat space, as we will see.
Note also that the matrix model action \eq{YM-action-1} 
can be written in the semi-classical limit as 
\be
S_{YM} = - \Lambda_0^4\, Tr \frac 14 [X^a,X^b][X^{a'},X^{b'}] \eta_{aa'} \eta_{bb'} 
\,\sim\, \frac 1{(2\pi)^2}\,\int d^{4} x\,  \Lambda_0^4\, \rho(x) \eta(x) ,
\label{S-semiclassical-general}
\ee
where 
\bea
\eta(x) &=& \frac 14 e^\sigma \, G^{\mu\nu}(x) g_{\mu\nu}(x) .
\label{eta-def}
\eea

\paragraph{Covariant equations of motion.}

As shown in \cite{Steinacker:2008ri}, 
the basic matrix equations of motion 
\eq{eom-vac-matr} can now be cast into a covariant form
as follows:
\bea
\Box_{G} \phi^i &=& 0  \label{eom-phi} \\
\Box_{G} x^\mu &=& 0 \label{eom-X-harmonic-tree}\\
\nabla_G^\eta (e^{\sigma} \theta^{-1}_{\eta\nu}) 
\, &=& \, e^{-\sigma}\,G_{\mu\nu}\,\theta^{\mu\g}\,\partial_\g\eta(x)
\label{eom-geom-covar-extra}
\eea
where we consider $x^\mu \sim X^\mu$ as a scalar function on
$\cM$, consistent with the ambiguity of the splitting  
$X^a = (X^\mu,\phi^i)$ into coordinates and scalar fields.
Here $\nabla_G$ denotes the Levi-Civita connection with respect to
the effective metric $G^{\mu\nu}$.
In particular, such on-shell geometries 
imply \cite{Steinacker:2008ri}
\be
\Gamma^\mu = 0 
\label{gamma-mu}
\ee
for the preferred matrix coordinates 
$x^\mu$, which in general relativity 
would be interpreted as gauge condition.
Furthermore, it turns out that \eq{eom-geom-covar-extra}
is in fact a consequence of a matrix Noether theorem 
due to the translational symmetry $X^a \to X^a + c^a \one$,
and is therefore protected from quantum corrections 
\cite{Steinacker:2008ya}.
It provides the relation between the noncommutativity 
$\theta^{\mu\nu}(x)$ and the metric $G^{\mu\nu}$.
Since \eq{eom-geom-covar-extra} has essentially the form
of covariant Maxwell equations coupled to an external
current, it should have a unique solution
for a given ``boundary condition''
\be
\theta_{\mu\nu}(x) \,\to
\, \bar \theta_{\mu\nu} = const 
\qquad \mbox{for}\quad |x|\to\infty 
\label{theta-asympt}
\ee
up to radiational contributions, which play the role of 
gravitational waves here \cite{Rivelles:2002ez,Steinacker:2007dq}.

\paragraph{Relation with string theory.}

The IKKT matrix model was proposed originally as a non-perturbative
definition of IIB string theory on $\R^{10}$. 
From this point of view, $\cM_\theta$ 
could be interpreted as a brane
with open string metric $G^{\mu\nu}$,
while $g_{\mu\nu}$ could be viewed as 
closed string metric in the 10D bulk. 
Indeed there are also other 
solutions of the matrix model, in particular 10D solutions. 
Notably graviton scattering has been studied 
in this and related matrix models; for an
incomplete list of references see e.g. 
\cite{Banks:1996vh,Banks:1996nn,Ishibashi:1996xs,Chepelev:1997ug,Aoki:1998vn,Aoki:1999vr,Nair:1998bp,Taylor:2001vb,Kitazawa:2006pj}
and references therein.
Most of this work was based on a different kind of
block-matrix backgrounds, and to simple geometries 
in the NC case. The essential point here is to consider
general 4-dimensional NC brane
solutions,  without physical 10D bulk. 
On such a background, the matrix model 
can be viewed as NC gauge theory 
\cite{Douglas:2001ba,Szabo:2001kg},
which includes gravity as we have seen. In this way
the strength of string theory  
(notably the good behavior under quantization) 
is preserved while the main problems
(lack of predictivity) are avoided.
In particular, the matrix model should be viewed as
background-independent here, since there is no 
physical space-time to start with.

A natural question arises how the present results relate 
to the BFSS matrix model \cite{Banks:1996vh} of (M)atrix theory, 
which involves 9 matrices depending 
on a classical ``time'' parameter.
This model is not unrelated to the IKKT model 
as pointed out in \cite{Ishibashi:1996xs}, 
and some version of the present mechanism should be
realized there as well; this should be studied elsewhere. We 
only want to stress here that 
the space-time brane $\cM$ must be entirely noncommutative
in the present framework, i.e. $\theta^{\mu\nu}$
must be non-degenerate. 
Nevertheless a conventional physical picture
emerges at scales larger than $\L_{NC}$, and 
there is no need for a classical time parameter. 
Hence the IKKT model is more natural for
the mechanism of emergent gravity.
It is also very interesting that some evidence for
4-dimensional space-time (rather than some other 
dimension) emerging in that model 
has been found \cite{Nishimura:2001sx}.

The topic of membranes and matrix models
has of course a long history, cf. also 
\cite{deWit:1988ig,Madore:1991bw,Nicolai:1998ic}.
It is also interesting to compare the present approach 
with other models of emergent gravity, see e.g.
\cite{Liberati:2009uq} and references therein. 


\subsection{Self-dual solutions and $g_{\mu\nu} = G_{\mu\nu}$}

A simple class of solutions of \eq{eom-geom-covar-extra} is given by 
``self-dual'' 2-forms $\theta^{-1}_{\mu\nu}(x)$ which 
satisfy
\be
\nabla^\mu \theta^{-1}_{\mu \nu} = 0, \qquad 
g_{\mu\nu} =  G_{\mu\nu} .
\label{self-dual-metric}
\ee
To see this, consider (at a point $x$) a local coordinate system where
$g_{\mu\nu} = \diag(s,1,1,1)$ (with $s=\pm 1$ in the
Euclidean resp. Minkowski case)
and $\theta^{\mu\nu}$ has the form 
\be
\sqrt{\rho}\, \theta^{\mu\nu} = \left(\begin{array}{cccc} 0 & 0 & 0 & -\sqrt{s}\a \\
                                0 & 0 & \pm\a^{-1} & 0 \\
                                0 & \mp\a^{-1} & 0 & 0  \\
                                \sqrt{s}\a & 0 & 0 & 0 \end{array}\right)\, ;
\label{theta-standard-general}
\ee 
this can always be achieved using a $SO(4)$ resp. $SO(1,3)$ 
transformation\footnote{Note that we assume that $\theta^{0i}$ 
is imaginary in the Minkowski case. This might appear strange,
but it is natural having in mind a Wick rotation $x^0 = i t$, 
and it is essential for $g_{\mu\nu} = G_{\mu\nu}$. 
This should be addressed in more detail elsewhere.} 
at the point $x$. Clearly this is (anti-) self-dual
$\star \theta^{-1} = \pm \sqrt{s}\,\theta^{-1}$  if and only if 
$\a^2 = 1$, where $\star$ denotes the Hodge star
and $\theta^{-1}= \theta^{-1}_{\mu \nu} dx^\mu\wedge dx^\nu$.
On the other hand, 
\be
G^{\mu\nu} = \rho\theta^{\mu\mu'} \theta^{\nu\nu'} g_{\mu'\nu'} 
= \diag(s \a^2,\a^{-2},\a^{-2},\a^2) 
\ee
at the point $x$,
hence $e^{-\sigma}\eta = \frac 12(\a^2+\a^{-2})$. Therefore
$\star \theta^{-1} = \pm \sqrt{s}\,\theta^{-1}$ if and only 
if $g_{\mu\nu} = G_{\mu\nu}$, or equivalently 
$e^{\sigma}= \eta$. 
It is then easy to see that 
in this self-dual case \eq{eom-geom-covar-extra}
reduces to $\nabla^\mu \theta^{-1}_{\mu \nu} = 0$, which holds
identically since $d\star\theta^{-1} = d\theta^{-1} = 0$. 

Such (anti-)self-dual closed 
2-forms $\theta^{-1}$ with constant asymptotics 
$\theta^{-1}_{\mu \nu}(x) \to \obar \theta^{-1}_{\mu \nu}$
as $x\to\infty$ always exist, at least on asymptotically flat spaces. 
This can be understood by interpreting 
$\theta^{-1} $ as 
sourceless electromagnetic field with constant field strength at infinity. 
Indeed, we only have to solve 
$d\star F = 0, \,\, F = dA$ with suitable asymptotics 
$F \to \obar F$ as $r \to\infty$, and 
define $\theta^{-1} $ to be
the (anti-)selfdual component of $F$.

In this self-dual case $g_{\mu\nu} = G_{\mu\nu}$,
the bare matrix model action \eq{S-semiclassical-general} becomes
\be
S_{MM} = \Lambda_{0}^4 \int d^4 x\, \rho \eta =  \Lambda_{0}^4 \int d^4 x\, \rho e^\sigma
= \int d^4 x\, \sqrt{|G_{\mu\nu}|}  \Lambda_{0}^4
= \int d^4 x\, \sqrt{|g_{\mu\nu}|}  \Lambda_{0}^4
\ee
which is precisely the form of the 
vacuum energy resp. brane tension, interpreted 
as cosmological constant in GR. 
We collect all such (bare and induced) terms in
\be
S_{vac} = - 2\int d^4 x\,\sqrt{|g_{\mu\nu}|} \Lambda_1^4
\label{vac-energy-action}
\ee
denoting with $\Lambda_1^4$ the sum of $\Lambda_0^4$ and the 
induced quantum-mechanical vacuum energy; 
the above sign is essential for stability reasons, 
cf. section \ref{sec:stability}. 
We also recall that the effective Yang-Mills coupling constant
for the basic $SU(n)$ gauge fields on $\cM^4$
is given by \cite{Steinacker:2008ya}
\be
g^2_{YM} = \Lambda_0^{-4}\, e^{-\sigma}  \sim \Lambda_0^{-4}\, \Lambda_{NC}^4 ,
\ee
so that $g_{YM} = O(1)$ amounts to 
\be
\Lambda_0 \sim \Lambda_{NC} .
\label{L0-LNC}
\ee
Let us briefly discuss deviations from
the self-dual case $g_{\mu\nu} \neq G_{\mu\nu}$. 
These arise from
(abelian) variations $\d x^\mu =  \theta^{\mu\nu} A_\nu$
of the tangential matrix degrees of freedom
resp. ``would-be $U(1)$ gauge fields'', which lead to
$\d\theta^{-1}_{\mu\nu} =  F_{\mu\nu}$
where $F_{\mu\nu}$ is the corresponding field strength.   
The corresponding 
propagating degrees of freedom behave as gravitational waves,
with metric fluctuations 
 $h_{\mu\nu}  =   G_{\nu \nu'} \theta^{\nu'\rho}  F_{\rho\mu} 
 +  G_{\mu \mu'} \theta^{\mu'\rho}  F_{\rho\nu} \, 
 + \frac 12 G_{\mu \nu} F_{\rho\eta} \theta^{\rho\eta}$ 
 \cite{Rivelles:2002ez,Steinacker:2007dq}.
In the presence of matter, we expect additional source terms 
on the rhs of equations \eq{eom-phi} -- \eq{eom-geom-covar-extra}.
While matter is not charged under this gravitational $U(1)$,
it is expected to induce e.g. dipole 
(and higher multipole) excitations of $F_{\mu\nu}$.
This would lead to short-distance modifications of the metric which 
decay more rapidly than $\frac 1r$. We therefore
simply ignore such effects here, and consider only the simplest
``self-dual'' solutions with $g_{\mu\nu} = G_{\mu\nu}$.

\subsection{Static metric deformations}

We focus  on deformations of the embedding 
$\cM^4\subset \R^D$ through
the scalar fields $\phi^i$ with embedding metric \eq{g-explicit}
\bea
g_{\mu\nu} &=& \eta_{\mu\nu} + \partial_\mu\phi^i\partial_\nu\phi^i \nn\\
&\equiv& \eta_{\mu\nu} + h_{\mu\nu} ,
\eea
and assume that $G_{\mu\nu} = g_{\mu\nu}$,
i.e. that $\theta^{\mu\nu}$ is self-dual w.r.t. $g_{\mu\nu}$
as discussed above.
This is an ansatz\footnote{In fact one 
of the main difficulties in this context is to understand and
disentangle the effects of the embedding $\phi^i$ versus 
the NC structure $\theta^{\mu\nu}$. We hope that this paper
helps to clarify this issue. In particular, the ansatz in 
\cite{Steinacker:2007dq} with trivial embedding but nontrivial 
$\theta^{\mu\nu}$ may not be realized in the presence
of matter.} which may not be completely appropriate,
but it is expected to give the correct long-distance physics
in the static case
since corrections due to $G_{\mu\nu} \neq g_{\mu\nu}$ 
lead to be short-distance effects and gravitational waves.

We furthermore focus on {\em static} metrics $g_{\mu\nu}$, 
corresponding to static 
and somewhat localized matter distributions.
One goal is to obtain the analog of the 
Schwarzschild solution at least in some regime.
Since $h_{\mu\nu}$ is quadratic in 
$\phi$,  we cannot restrict ourselves to linearized
fluctuations in $\phi$ around flat Minkowski space.
We must include at least quadratic terms in $\phi$,
which makes the analysis somewhat non-trivial. 
Nevertheless we will find a class of solutions which 
appear to give the appropriate
description of objects such as galaxies sparsely 
embedded in flat Minkowski space. 
This also fits naturally 
in the context of the cosmological solutions \cite{Klammer:2009ku}, 
as discussed in section \ref{sec:cosmology}.

Some insight can be gained by observing that the metric fluctuation
$h_{\mu\nu}$ is closely related to the energy-momentum tensor of the 
massless scalar fields $\phi^i$,
\bea
T_{\mu\nu}^\phi &=& \partial_\mu\phi \partial_\nu\phi 
- \frac 12 \eta_{\mu\nu} (\eta^{\a\b}\partial_\a\phi \partial_\b\phi)
= h_{\mu\nu} - \frac 12 \eta_{\mu\nu} \, h,  \nn\\
h  &=& \eta^{\a\b} h_{\a\b} = - \eta^{\mu\nu} T_{\mu\nu}^\phi .
\eea
In particular, the effective Newtonian potential is
related to the  energy density of a free scalar field,
\be
2U(x) = -h_{00} = -T_{00}^\phi + \frac 12 \, h 
\approx -\frac 12 T_{00}^\phi
\ee
assuming $T_{ij}^\phi \approx 0$ in the last step.
Hence a static metric fluctuation corresponds to a 
a scalar field excitation with static energy-momentum 
tensor.
Furthermore, 
since $\phi$ in vacuum satisfies the free wave equation 
\eq{eom-phi}, it will decay like $\frac 1r$
outside of matter distributions, suggesting a 
$U(r) \sim \frac 1{r^2}$ behavior for the gravitational potential.
This appears very bad at first, since this is not
the usual $U(r) \sim \frac 1r$ law of Newtonian gravity. 

The resolution of this puzzle is as follows: 
the relevant {\em non-singular} harmonic
excitations  necessarily have some finite wavelength,
$\phi^i \sim \frac{\sin(\omega r)}{\omega r} e^{i\omega t}$.
These are localized excitations resp. ``gravity bags'', 
which we will argue to have 
very long (astronomical) wavelength $\omega$. This leads to 
a long-distance screening of gravity with $U(r) \sim \frac 1{r^2}$.
However the crucial point is that {\em inside} these gravity bags
(in particular inside of galaxies, say), the local
matter distribution does indeed lead to the standard 
$U(r) \sim \frac 1{r}$, as we will show. 
Newtonian gravity therefore arises as a
``short-distance'' effect on the 
harmonically embedded spacetime brane.

\paragraph{Basic ``rotating'' embedding.}

We consider brane embedding $\cM^4\subset \R^D$
which are small deformations of flat Minkowski space, 
with the following structure:
\bea
x^A = \(\begin{array}{c} x^0 \\ x^i \\ \phi^i\end{array}\)
= \(\begin{array}{c} t  \\ x^i \\ 
g(x)\, \(\begin{array}{c}\cos(\omega t) \\ \sin(\omega
  t)\end{array}\)\\
\end{array}\)
\label{basic-radial-embedding}
\eea
where $g(x)$ is independent of $t$ and $g(x) \to 0$ as $r \to \infty$.
There might be additional components in this ansatz.
We stress that the $\phi^i$ comprise {\em two space-like components}  
which {\em rotate} 
with $t$, rather than just of its components\footnote{The 
time-like component(s) could have 
similar nontrivial  embeddings. This should be 
elaborated elsewhere.}.
We will see in  section \ref{sec:cosmology}
that due to its flat asymptotics,
this type of embedding fits naturally into the context of the 
cosmological solutions obtained in \cite{Klammer:2009ku}.
An important feature of this ansatz is 
\be
\partial_0 \phi^k \partial_i \phi^k = 0,
\label{0i-component}
\ee
leading to a {\em static}  metric 
\be
d s^2 = -(1- \omega^2 \, g^2) \, dt^2 + 
(\d_{ij} + \partial_i g \partial_j g) dx^i dx^j 
\label{basic-metric-static}
\ee
This is consistent with the fact that
there is no energy flux associated to standing waves.
There might be several components of $g \to g^i$
whose contributions will simply add up; this will not alter the
essential conclusions below.
Clearly the rotating embedding with $\omega \neq 0$ must
be responsible for Newtonian gravity, which cannot
be reproduced with a purely static embedding.
Additional components might be necessary 
to comply e.g. with the detailed solar system constraints. 
The embedding functions $\phi^i$ must be determined by 
solving the equations of motion in the presence of matter,
which will be done below.
Note that the preferred ``matrix coordinates'' $x^\mu$
automatically satisfy the e.o.m. $\Box_g x^\mu =0$ 
in vacuum \eq{eom-phi}, \eq{eom-X-harmonic-tree};
therefore we will only consider the equations of motion for 
$\phi^i$ below.

\section{Coupling to matter and effective gravity}

\subsection{Effective action and equations of motion with matter}

In this section we derive the equations of motion for 
gravity coupled to matter in the semi-classical 
limit, assuming $g_{\mu\nu} = G_{\mu\nu}$
as discussed above. Our starting point is the
semi-classical effective action of the matrix model 
\eq{vac-energy-action} together with the action for matter,
and the Einstein-Hilbert 
action $R[G] = R[g]$ which is 
induced at one-loop\footnote{Assuming
a cutoff $\Lambda_4$, which is the cutoff for $N=4$ SUSY in the 
IKKT model, see \cite{Grosse:2008xr,Matusis:2000jf}} 
\be
S = \int d^4 x \sqrt{|g|}\, (\Lambda_4^2 R - 2\Lambda_1^4) + S_{\rm matter} .
\label{basic-action}
\ee
Now recall
\bea
\d \int\sqrt{|g|} &=&  \d \sqrt{-\det g} = \frac 12 \sqrt{|g|} \, g^{\mu\nu} \d g_{\mu\nu},  \nn\\
\d \int\sqrt{|g|}\,  R &=& - \sqrt{|g|}\,\cG^{\mu\nu} \d g_{\mu\nu} \nn\\
\d S_{\rm matter} &=& 8\pi \sqrt{|g|}\,T^{\mu\nu} \d g_{\mu\nu} 
\eea
where $\cG^{\mu\nu} = R^{\mu\nu}- \frac 12 g^{\mu\nu} R$ 
is the Einstein tensor, 
and $T^{\mu\nu} $ is the energy-momentum tensor
for matter 
(recall that matter and fields couple to the effective metric 
essentially in the standard way). 
The crucial point is now that the fundamental
geometrical  degrees 
of freedom are not the $g_{\mu\nu}$, 
but the embedding fields $\phi^i$ as well as $X^\mu$ 
resp. $\theta^{\mu\nu}$. In the self-dual case 
$g_{\mu\nu} =G_{\mu\nu}$, 
the most general variation 
can thus be decomposed into variations of 
$g_{\mu\nu} = \eta_{\mu\nu} + \partial_\mu\phi^i \partial_\nu\phi^i 
= G_{\mu\nu}$
and variations of $\theta^{\mu\nu}$. 
The e.o.m. for $\theta^{\mu\nu}$
are satisfied\footnote{This however amounts
to neglecting matter contributions to $\theta^{\mu\nu}$,
as discussed before.
For geometries with non-selfdual $\theta^{\mu\nu}$
this derivation must be refined.} 
due to \eq{self-dual-metric}.
The variation
of $S$ with respect to the fundamental fields $\phi^i$ 
can be written using $\d g_{\mu\nu} = \partial_\mu \phi^i \partial_\nu\d \phi^i  
+ \partial_\mu \d\phi^i \partial_\nu\phi^i $ as
\be
\d S = \int d^4 x \sqrt{|g|}\, 
\d g_{\mu\nu} \cH^{\mu\nu} 
= -2\int \d \phi^i \partial_\mu(\sqrt{|g|}\,\cH^{\mu\nu})\partial_\nu\phi^i
\label{variation-cutoffs}
\ee
up to boundary terms, where 
\be
\cH^{\mu\nu} 
=  8\pi T^{\mu\nu}- \Lambda_4^2\cG^{\mu\nu} - \Lambda_1^4 g^{\mu\nu} .
\label{H-def}
\ee
This  leads to the equations of motion for $\phi^i$
\bea
\partial_\mu(\sqrt{|g|}\, \cH^{\mu\nu}\partial_\nu\phi) = 0 ,
\eea
which using the identity $\nabla_\mu V^{\mu} \equiv
\frac 1{\sqrt{g}}\partial_\mu(\sqrt{g}\, V^{\mu})$
 can be written as
\bea
 \Lambda_1^4 \Box_g \phi
&=& (8\pi T^{\mu\nu} - \Lambda_4^2\cG^{\mu\nu})\nabla_\mu\partial_\nu\phi
+ 8\pi (\nabla_\mu T^{\mu\nu})\partial_\nu\phi
\label{eom-general}
\eea
recalling  $\nabla_\mu \cG^{\mu\nu} =0$.
This equation has 2 types (``branches'') of solutions: 
\begin{enumerate}
\item
\underline{``Einstein branch''}: 

Clearly every solution of the Einstein equations 
\be 
\Lambda g^{\mu\nu} + \cG^{\mu\nu} = 8\pi G\, T^{\mu\nu}
\label{E-H-solution}
\ee
is also a solution of $\cH^{\mu\nu} =0$ \eq{eom-general}, upon
identifying
\be
\frac {1}{G} =  \Lambda_4^2 \qquad
\frac {\Lambda}{G} = \Lambda_1^4 .
\label{E-H-indentification}
\ee
Indeed using general embedding theorems \cite{clarke}, 
any solution of the E-H equations can be realized (locally)
using embeddings in $D\geq 10$. 
This should  provide an interesting realization
of Einstein gravity within matrix models, with
technical advantages over other approaches in particular 
for quantization. 
Recall however that we  assumed $G_{\mu\nu} = g_{\mu\nu}$,
which will not always hold; thus there will be some modifications 
of the Einstein equations due to $\theta^{\mu\nu}(x)$.
On the other hand, this Einstein branch would presumably lead to the
same problematic aspects of GR, notably fine-tuning issues
in cosmology. 
In this paper we will focus on the second kind of solution,
which leads to significant and very interesting deviations 
from GR at long distances.

\item
\underline{``Harmonic branch''}: 

Besides to the above solutions, there are
additional solutions of \eq{eom-general} with
\be 
\partial_\mu(\sqrt{|g|}\, \cH^{\mu\nu}\partial_\nu\phi) = 0 , \qquad
\cH^{\mu\nu}  \neq 0
\label{harmonic-solution}
\ee
The prototype of such a solution 
without matter is flat Minkowski space $\phi^i=0$ with $\Lambda_1>0$.
More generally if the vacuum energy dominates the matter density,
then \eq{eom-general} reduces to $\Box_g \phi \approx 0$,
leading essentially to minimal surfaces
which will be deformed in the presence of matter.
This in turn leads to the near-realistic cosmological 
solutions of FRW type 
found in \cite{Klammer:2009ku} with Milne-like
late-time behavior, which are stable and largely 
insensitive to the detailed matter content. 
Another attractive feature of this harmonic 
branch is that its quantization should be comparably
simple, as the embedding fields $\phi^i$ are governed
in vacuum by a simple action with positive excitation spectrum,
cf. section \ref{sec:stability}.

\end{enumerate}

We will simply assume in the following that 
$\Lambda_2^4\cG^{ij}$ can be neglected in the static case. 
This is however far from trivial, since the
simple ansatz \eq{basic-radial-embedding}
will generally {\em not} lead to $\cG^{ij}\approx 0$. 
There are 2 ways how this might nevertheless be justified: either
a more sophisticated ansatz taking into account $\theta^{\mu\nu}$
will imply $\cG^{ij}\approx 0$, or $\Lambda_4^2$ is much smaller
than the Planck scale. The latter is in fact very appealing
as we will see. In any case we focus on the  harmonic branch 
in this paper, since it leads to 
very interesting long-distance modifications 
of Newtonian gravity which is the
main topic of this paper.

\subsection{Perturbations of flat Minkowski space-time}
\label{sec:perturbations}

For simplicity we focus on perturbations of flat Minkowski space,
and only consider the case of ``weak'' gravity, i.e.
\be
g_{\mu\nu} = \eta_{\mu\nu} + h_{\mu\nu}
\ee
keeping only terms linear in 
$h_{\mu\nu} = \partial_\mu \phi^i \partial_\nu \phi^i$. 
The e.o.m. \eq{eom-general} then simplifies to 
\be
 \Box_\eta \phi =  \frac{8\pi }{\Lambda_1^4}\, \tilde T^{\mu\nu}
\partial_\mu\partial_\nu\phi , \qquad
\tilde T^{\mu\nu} \equiv T^{\mu\nu} - \frac{\Lambda_4^2}{8\pi } \cG^{\mu\nu}
\label{eom-general-linear}
\ee
using\footnote{The conservation law for $T^{\mu\nu}$
is not entirely evident in the matrix model, in particular for 
fermions which have a non-standard spin connection \cite{Klammer:2008df}.
Nevertheless it is expected to hold 
at least to a very good approximation since the matter action almost 
coincides with the usual one.} $\nabla_\mu T^{\mu\nu}=0$. 
We also replace $\Box_g \approx \Box_\eta$ 
to leading approximation.
In the case of a static mass distribution $\rho$, 
we assume furthermore that 
\be
\tilde T_{00} = \rho(x) \equiv \rho^{matter}(x) - \rho^{curv}(x) \geq 0, 
\qquad 
\rho_{curv} = \frac{\Lambda_4^2}{8\pi } \cG^{00}
\ee
and $\tilde T_{ij} \approx 0$ as discussed above\footnote{One 
might think that due to the Einstein equations, 
$\rho = \rho_{matter} - \rho_{grav} = 0$ would drop
out of the harmonic equation. This is not the case here, as the 
Einstein equations do not hold in this form.}. 
Then \eq{eom-general-linear} 
can be written using $\Box = - \partial_t^2 + \Delta$ as
\be
\Delta \phi =  (1+\frac{8\pi}{\Lambda_1^4}\rho(x)) \partial_0^2\phi . 
\label{eom-general-linear-rho}
\ee
Note also that
\bea
\partial^\mu h_{\mu\nu} &=& (\partial_\mu\phi)(\partial^\mu \partial_\nu\phi) 
+ (\Box \phi)(\partial_\nu\phi) \nn\\
 &=& \frac 12 \partial_\nu h 
+ (\frac {8\pi }{\Lambda_1^4} \tilde T^{\a\b}\partial_\a\partial_\b\phi)(\partial_\nu\phi)
\label{h-gauge}
\eea
so that $h_{\mu\nu}$
satisfies the ``harmonic gauge''
$\partial^\mu h_{\mu\nu}= \frac 12 \partial_\nu h$
up to corrections due to matter and curvature.

\subsubsection{Harmonic gravity bags}

We are mainly interested in static spacetime-geometries 
in this paper. Thus
consider the following localized excitation of the 
embedding \eq{basic-radial-embedding}
\be
\phi^i(x,t) = g(x) e^{i\omega t} 
= g(x) \(\begin{array}{c}\cos(\omega t) \\ \sin(\omega t)\end{array}\)
\label{harmonic-ansatz}
\ee 
with some very small $\omega$, leading to the
 static  metric \eq{basic-metric-static}
\be
d s^2 = -(1- \omega^2 \, g^2) \, dt^2 + 
(\d_{ij} + \partial_i g \partial_j g) dx^i dx^j .
\ee
For $\rho = 0$ and neglecting 
the curvature corrections $\Lambda_4^2\cG_{\mu\nu}$,
the equation of motion \eq{eom-general-linear-rho} becomes
\be
\Box \phi_0(x) = 0,  \qquad 
\Delta g_0(x) = -\omega^2 g_0(x) .
\ee
Consider first the case of a spherical wave, where
this equation reduces to
\be
\partial_r(r^2 g') + \omega^2 r^{2} g(r) = 0 .
\label{g-equation-linear}
\ee
The unique spherically symmetric solution which is
regular at the origin and decays as $r \to \infty$ is given by 
\bea
g_0(r) &=& g_0 \frac{\sin(\omega r)}{\omega r} , \nn\\
\phi_0(x) &=& g_0(r)\, e^{i\omega t}
\, = \, g_0(r)\, 
\(\begin{array}{c}\cos(\omega t) \\ \sin(\omega t)\end{array}\)
\label{phi-0}
\eea
with radial wavelength given by 
\be
L_\omega = \frac{\pi}{\omega} .
\label{L-omega}
\ee
Near the origin, we can write
\bea
g_0(r) &=& g_0 \frac{\sin(\omega r)}{\omega r} \approx 
g_0 (1-\frac 16 \omega^2 r^2 + ...) , \nn\\
g'(r) &=& g_0  \Big(\frac{\cos(\omega r)}{r} - \frac{\sin(\omega
  r)}{\omega r^2}\Big)
\approx g_0 (-\frac 13 \omega^2 r + \frac 1{30} \omega^4 r^3 + O(r^5)).
\eea
Note that  $g'$ is regular at the origin, and decays like
$\frac 1{r}$ as $r \to \infty$.
The effective metric \eq{basic-metric-static} is
\be
d s^2 = -(1- \omega^2 \, g(r)^2) \, dt^2 + (1+(g')^2) dr^2 
+ r^2 d\Omega^2 ,
\ee
which allows to read off the effective
gravitational potential: a static test particle 
in this metric perturbation feels an effective gravitational potential 
\bea
g_{00} = -(1+2U_0), \qquad 
2U_0(r) &=& - \omega^2 \, g(r)^2 =  -\omega^2 g_0^2 
\Big(\frac{\sin(\omega r)}{\omega r}\Big)^2 , 
\eea
which satisfies
\bea
U_0(r) &\sim& - \omega^2 \frac 1{r^2}, \qquad r \to \infty 
\label{U-vac-long} \\
\Delta U_0(r) &=&  g_0^2 \omega^4 , \qquad\quad r \sim 0
\label{Delta-U-vac}
\eea
The last equation will be interpreted later
in terms of an effective 
gravitational constant/vacuum energy $\Lambda_{\rm eff}$ \eq{Lambda-eff}
inside the bag. For large distances, 
$U(r)$ leads to
a rapidly decreasing attractive gravitational force
with range $L_\omega$. This might appear strange at first, 
since such vacuum excitations do not exist in general relativity.
However, they are not expected to survive in this naked form:
Due to their attractive 
gravitational force, matter will tend to accumulate inside, and
these ``gravity bags'' will typically 
enclose matter with $\rho\neq 0$. In particular,
large clusters of matter such as galaxies will  
be embedded in such ``gravity bags''.
The essential point is that the matter {\em within} such a 
gravity bag will experience Newtonian gravity,
as we will show next.

\subsubsection{Spherically symmetric mass distribution and Newtonian limit}

We now show how Newtonian gravity arises in the 
simplest case of spherically symmetric localized mass
distributions, due to a local deformation of the
above harmonic gravity bag. This will lead moreover to 
significant long-distance deviation from Newtonian gravity and GR. 
Basically, gravity is confined
within the bag, which also contains an effective vacuum energy. 
The case of a more general mass distribution 
will be studied in section \ref{sec:general-rho}. This will 
clarify the significance of the gravity bag which 
we simply assume here.

Consider a spherically symmetric mass distribution
around the origin within the radius $r_M$. For $r>r_M$,
the corresponding solution of \eq{eom-general-linear-rho}
with the ``static'' ansatz \eq{harmonic-ansatz}
must have the form 
\bea
\phi^i &=& g(r) e^{i \omega t} , \nn\\
g(r) &=&  g_0 \frac{\sin(\omega r+\d)}{\omega r} 
\quad   \sim \,\,  g_0 (\cos(\d) + \frac{\sin(\d)}{\omega r}) 
\label{g-r-phase}
\eea 
assuming $\omega r\ll 1$ in the last expression.
This phase shift $\d\neq 0$ 
is the key for obtaining Newtonian gravity.
It leads to the effective metric
\bea
g_{00} &=& -(1-\omega^2 g^2) \nn\\
&=& -(1 - g_0^2 \omega^2\cos(2 \d)) + \frac{g_0^2 \omega\sin(2\d)}{r} 
+\frac{g_0^2 \sin^2(\d)}{r^2} 
 - \frac 23 g_0^2 \omega^3 \sin(2\d)\, r \nn\\
&& - \frac 13 g_0^2 \omega^4 \cos(2\d) r^2 
 + O(r^3) , \label{g00-simple}\\
g_{rr} &=& 1+ g'^2  \nn\\
&=& 1+ \frac 13 \frac{g_0^2 \omega\sin(2\d)}{r}
 + \frac{g_0^2 \sin^2(\d)}{r^2} 
 + \frac{g_0^2 \sin^2(\d)}{\omega^2 r^4} 
 + \frac{2}{15} g_0^2 \omega^3 \sin(2\d) r \nn\\
&& + \frac 19 g_0^2 \omega^4 \cos(2 \d) r^2 + O(r^3) . 
\label{grr-simple}
\eea
This corresponds to a gravitational potential 
$U(r) = - \frac 12 \omega^2 g^2$, which for intermediate distances
\be 
\sin\d \ll \omega r \ll 1
\ee 
is well approximated by a $\frac 1r$ potential with a constant shift,
\be
U(r) = - \frac 12 \omega^2 g^2 
\approx \frac 12 g_0^2 \omega^2\cos(2 \d) -\frac{g_0^2 \omega\sin(2\d)}{2r} .
\ee
Thus all we need to obtain Newtonian gravity is $\d\sim M$,
which is very intuitive and indeed correct as shown below.

\paragraph{Boundary condition.}

Now consider the region near the origin where $\rho(r) \geq 0$.  
Equation \eq{eom-general-linear-rho} gives
\be
0 = \omega^2 r^{2} (1+\frac{8\pi}{\Lambda_1^4}\rho) g(r) + \partial_r(r^2 g').
\ee
We focus on the region near the origin, where\footnote{this
will be justified in section \ref{sec:general-rho}, noting that 
$g_0$ is due to a 
large-scale background structure which
dominates the local perturbation due to $M$.} $g \approx g(0)$.
Then  
\bea
r^2 g'(r) &=& -\omega^2 \int_0^r d r' g(r) r'^{2} (1+\frac{8\pi}{\Lambda_1^4}\rho) \nn\\
&\approx& -\omega^2 g(0) \int_0^r d r' r'^{2} (1+\frac{8\pi }{\Lambda_1^4}\rho)  \nn\\
&\approx& -\omega^2 g(0) \(\frac{r^{3}}{3} + \frac{2M(r)}{\Lambda_1^4}\)  
\eea
hence
\be
\frac{g'(r)}{g(0)} \approx -\omega^2 \(\frac{r}{3} + \frac{ 2M(r)}{r^2 \Lambda_1^4}\)
\label{g-integrated}
\ee
where $M(r)$ is the mass inside the sphere with radius $r$ (including
possibly curvature contributions).
Assuming that $\rho(r) = 0$ for $r>r_M$,
we can match this with \eq{g-r-phase} which is valid
for $r\geq r_M$:
\bea
g(r) &=&  g_0 \frac{\sin(\omega r+\d)}{r}  \nn\\
\frac{g'}{g_0} &=& \cos(\d) (\omega \frac{\cos(\omega r)}r 
- \frac{ \sin(\omega r)}{r^2}) 
- \sin(\d)(\frac{\cos(\omega r) }{r^2} +\omega \frac{\sin(\omega r)}r) \nn\\
&\approx&  - \frac{\sin(\d)}{r^2} (1 + \frac{\omega^2 r^2}{2}) 
- \frac 13 \omega^3\cos(\d) \, r + O(\omega^4 r^2) 
\eea
assuming $\omega r \ll 1$.
Note that the singular terms are misleading, since this
expression is valid only for $r>r_M$. 
Combining this with \eq{g-integrated} gives the matching condition 
\bea
\frac{g'(r)}{g_0} 
\approx \frac{\sin(\d)}{\omega r^2} + \frac 13 \omega^2 \cos(\d) r 
&\stackrel{!}{=}& \omega^2 \frac{g(0)}{g_0}
\(\frac{r}{3} + \frac{2M}{r^2 \Lambda_1^4}\) 
\nn
\eea
where we neglect the constant term since $\omega r \ll 1$. 
This implies the two  conditions 
\bea
\frac{g(0)}{g_0}\frac{2M}{\Lambda_1^4} \omega^3 &=& \sin(\d), \nn\\
\frac{g(0)}{g_0} &=& \cos(\d) .
\label{delta-match}
\eea
Thus the time-component of the effective metric \eq{g00-simple} is
\bea
g_{00} &\approx&  - 1 + g_0^2 \omega^2 
+ 4 \frac{g_0^2\omega^4}{\Lambda_1^4} \frac{M}{r} 
 - \frac 43 \frac{g_0^2 \omega^4}{\Lambda_1^4} M\, \omega^2  r 
- \frac 13 g_0^2 \omega^4 r^2 \,
 + \frac{4M^2 g_0^2 \omega^6}{\Lambda_1^8 r^2} + O(r^3)  \nn\\
&\equiv& - \(1 +2 U_0 - 2 \frac{G M}{r} 
 + \frac 23 M G \,\omega^2 r - \frac 13 \Lambda_{\rm eff} r^2
+ (\frac{M G}r)^2 \frac 1{2U_0} \)\,\, + O(r^3)
\label{metric-spherical}
\eea
where we define
\bea
U_{0} &=&  - \frac 12  g_0^2 \omega^2,\label{U-const} \\
G &=& \frac{2g_0^2\omega^4}{\Lambda_1^4} 
= - 4 U_0 \,\frac{\omega^2}{\Lambda_1^4}
\label{Newton-constant} \\
\Lambda_{\rm eff} &=& - \frac 12 G \Lambda_1^4 = 2 U_0 \omega^2,
\label{Lambda-eff}
\eea
(note that $G$ is naturally small given that $L_\omega$ is large),
assuming $\d \ll 1$ and $\omega r \ll 1$.

Let us discuss the importance of the various terms.
The $O(M^2/r^2)$ term 
\be
\frac{g_0^2 \sin^2(\d)}{r^2} 
 = (\frac{M G}r)^2 \frac 1{2U_0} ,
\ee 
can be neglected compared with the Newtonian
term provided 
\be
\sin \d < \omega r \quad\mbox{i.e.}\quad 
 \frac{M G}{r} < g_0^2 \omega^2  = |2U_0| ,
\label{outside-condition}
\ee
which we assume for simplicity. 
This means that the Newtonian potential due to the 
local mass $M$ should be smaller
than the background potential $U_0$ due to the 
harmonic bag\footnote{Having in mind e.g. a star within a galaxy; 
this will become more obvious in section 
\ref{sec:general-rho} in the context of a general mass distribution.}.
This also implies that the vacuum energy term
(as well as the Newtonian potential) dominates the 
linear term,  
\bea
|\Lambda_{\rm eff} r^2| = 2 |U_0| \omega^2 r^2 \gg 
M G \,\omega^2 r = \frac{M G}r \omega^2 r^2 
\eea
which we will omit henceforth. Thus
the time-component $g_{00}$ 
of the effective metric \eq{metric-spherical} has 
approximately the form of a Schwarzschild-de Sitter 
metric with a constant shift \cite{Rindler:2006km},
\be \large{
\fbox{$\,\,
g_{00} \approx - \Large(1 +2 U_0 - \frac{2G M}{r} 
 - \frac 13 \Lambda_{\rm eff} r^2 \Large) \,\, $}}
\label{g00-approx}
\ee
assuming $\frac{MG}{U_0} < r < L_\omega$ 
and dropping the linear term $\frac 23 \frac{G M}{r} \omega^2r^2$. 
The Newtonian term  dominates the $\Lambda_{\rm eff}r^2$ term 
provided the vacuum energy 
\be
E_{\rm vac}(r) := \frac{4\pi r^3}{3} \Lambda_1^4 \ll M ,
\label{E-vac-1}
\ee (cf. \eq{E-vac-2}) inside $r$ is smaller than the mass $M$.   
We then obtain Newtonian gravity with
potential \be U(r) \,\, \approx \,\, U_0 - \frac{G M}{r} .
\label{U-Newton-r} 
\ee
This will be generalized to the case of an arbitrary mass distribution
in section \ref{sec:general-rho}.
Finally, we note that
\be
U_{0} 
= -\frac{3}{16\pi^3} \frac{G E_{\rm vac}(L_\omega)}{L_\omega}
\label{U-const-2}
\ee
can be interpreted as Newtonian potential due to the vacuum energy
contained within the harmonic bag of size $L_\omega$.

The basic result is that 
Newtonian gravity arises at intermediate scales, with
important long-distance modifications.
Notice that the precise form of the induced gravitational
action was never used up to now, rather
gravity arises through a deformation of the 
harmonic embedding  which couples to $T^{\mu\nu}$.
Thus the mechanism is quite different from GR.

\subsubsection{Deviations from Newtonian gravity}

We have shown that under the above assumptions, $g_{00}$ has 
essentially the form of a Schwarzschild-de Sitter 
metric with an {\em apparent negative} 
cosmological constant $\Lambda_{\rm eff}$  \eq{Lambda-eff}
which is related to the vacuum energy. 
However, note that the sign of $\Lambda_{\rm eff}$ is 
{\em different from GR} \eq{E-H-indentification}. 
This is not a mistake, but underscores the fact that
the physics of vacuum energy is different here from GR.
Inside the gravity bag, the
vacuum energy $\L_1^4 > 0$ contributes a positive energy density to 
the gravitational potential.
Note also that \eq{Delta-U-vac} for the harmonic gravity bag
can now be written as\footnote{recall that in GR, 
a negative $\L$ leads to an additional attractive gravitational
field as above, due to 
$\Delta U = 4\pi G \rho_{\rm matter} - \Lambda$, see \cite{Rindler:2006km}.}
\be
\Delta U_0(r) =  -\Lambda_{\rm eff} =  4\pi G\, \frac{\Lambda_1^4}{8\pi}
\ee
corresponding to an effective energy density due to $\Lambda_1^4$.
However 
for very large distances $r \geq L_\omega$, the effective metric 
approaches the harmonic behavior
\be
U(r) \sim - \frac 12 \omega^2 g_0^2 \frac{\sin^2(\omega r)}{r^2}
\ee
which is rapidly decaying and oscillating, smoothly merging with the 
flat large-scale metric $g_{\mu\nu}(x) \to \eta_{\mu\nu}$
resp. the Milne-like cosmology \cite{Klammer:2009ku}
as discussed in section \ref{sec:cosmology}.
In particular, we will see that 
cosmology does {\em not} lead to the usual 
stringent constraints on the vacuum energy.
However we obtain an upper bound for $\Lambda_1$ e.g. due to
solar system constraints (for example $\Lambda_1 = O(eV)$ would work, but
certainly not $\Lambda_1 = O(TeV)$).

\begin{figure}
  \vspace{-0.3cm}
\begin{center} 
\includegraphics[scale=0.2]{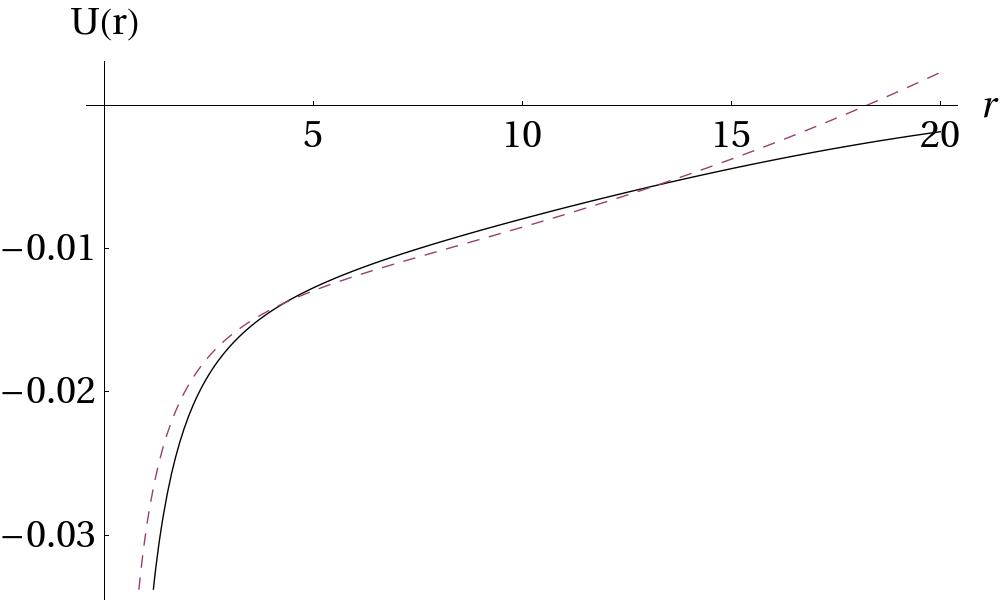}\\
\includegraphics[scale=0.2]{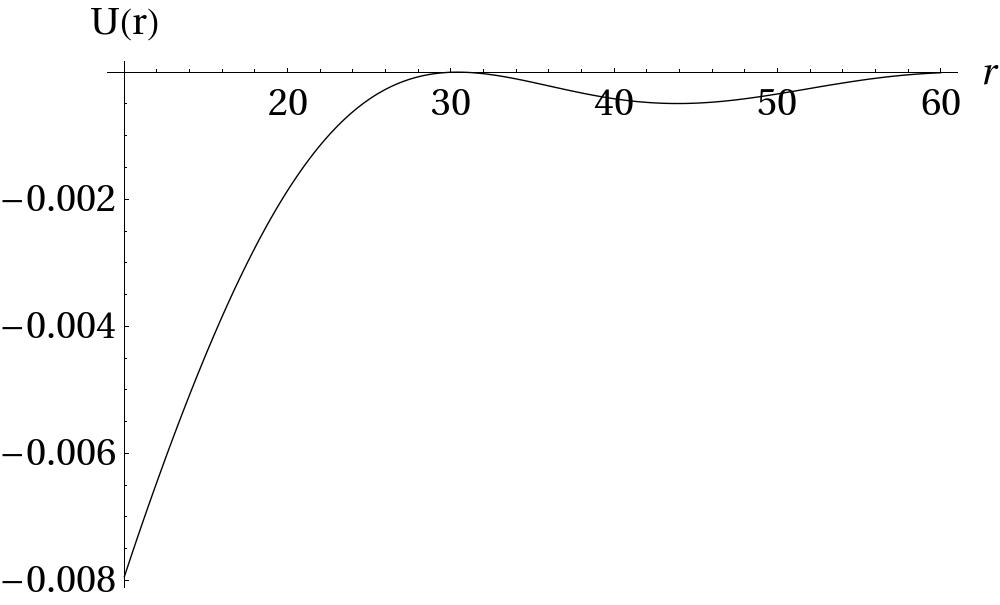}
\end{center}
\caption{Gravitational potential $U(x)$  compared with the
Schwarzschild-de  Sitter potential 
$U_0 - \frac{MG}{r} - \frac 16  \Lambda r^2$ 
(dashed line) and long-distance oscillations, for
$\omega = 0.1, \, g_0 = 1, \d = 0.1$.} 
\label{fig:U}
\end{figure}

A plot of the full $U(x) = -\frac 12 \omega^2 g(x)^2$ 
in comparison with the terms in \eq{g00-approx}
is given in figure \ref{fig:U} for
$\omega = 0.1, \, g_0 = 1, \d = 0.1$ hence $U_0 = - 0.005$.
This clearly shows the dominating Newtonian
form for small $r$. The vacuum energy term 
takes over for larger  $r$, until the potential is cut off 
effectively at $L_\omega$. 
Note that the amplitude of these oscillations 
is very small and 
thus may be very hard to detect, although in principle
such oscillations should be present at very large distance 
from isolated spherical galaxies. They might be masked for various
reasons such as matter distributions or
the superposition of slightly different $\omega_i$.

Now consider the radial part of the effective metric 
\eq{grr-simple}, which can be written as
\bea
g_{rr}
&\approx& 1+ \frac 13 \frac{2GM}{r} 
- \frac 19  \Lambda_{\rm eff} r^2\,\, 
 - \frac{GM}{r} \frac{1}{2\Lambda_{\rm eff} r^2} \frac{GM}{r}(1+\omega^2 r^2) 
 \, + O(r^3) 
\label{grr-simple-2}
\eea
There is a strange factor $\frac 13$ in 
\eq{grr-simple-2} compared with general
relativity. 
However, one should keep in mind that this metric is appropriate 
only for isolated masses resp. galaxies, but not for small 
perturbations within galaxies such as the solar system. 
The latter will be 
studied in section \ref{sec:general-rho}, confirming 
the factor $\frac 13$ for stars within galaxies, while the 
more singular terms are smaller in that case.
This may be a challenge
for the solar system constraints. However, 
there will be short-distance 
corrections e.g. due to $\theta^{\mu\nu}(x)$, and a more complete
analysis is required before a reliable judgment can be 
given. 

Note incidentally that the metric is regular but becomes 
degenerate as $g_{00} \to 0$.  
However, then the approximation of linearized gravity 
in \eq{eom-general-linear} is no longer valid, 
and a more careful treatment is required.

\paragraph{(Galactic) rotation curves.}

\begin{figure}
  \vspace{-0.2cm}
\begin{center}
 \includegraphics[scale=0.22]{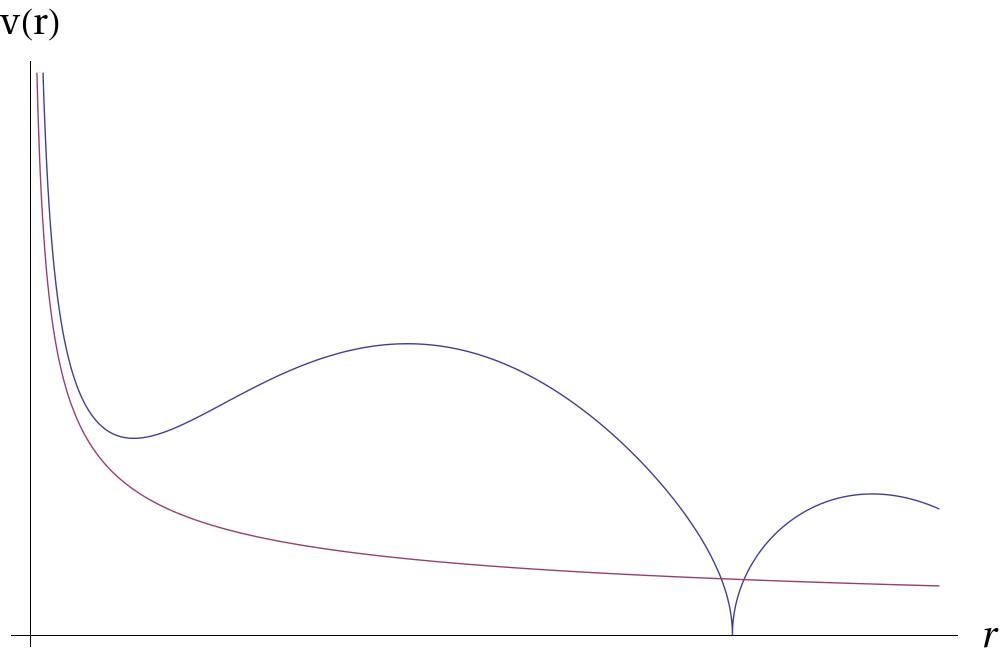}
\end{center}
\caption{Orbital velocity $v(x)$ for a central point mass
compared with Newtonian case:
Newtonian domain, enhancement and cutoff, for
$\omega = 0.1, \, g_0 = 1, \d = 0.1$.}
\label{fig:V}
\end{figure}

Now consider (non-relativistic) orbital velocities.
For small distances  $r < L_\omega$, it is given by
\be
v = \sqrt{U' r} = \sqrt{2 \frac{G M}{r} 
(1+\frac{\pi^2}{3} \frac{r^2}{L_\omega^2})
 - \frac 23 \Lambda_{eff} r^2} .
\label{v-SdS}
\ee
This decreases like $r^{-1/2}$ as in Newtonian
gravity as long as $E_{vac}(r) < M$,  
but for $E_{vac}(r) \approx 4\pi M$ it starts to increase linearly
like $v \sim \sqrt{|\Lambda_{eff}|}\, r$
until $r \approx L_\omega$; recall that $\Lambda_{eff}<0$. 
At that scale, the velocities 
decrease again with oscillating behavior and
\be
v = \sqrt{U' r} \approx \frac{\sqrt{2}}{r} g_0 
\label{harmonic-decay-v}
\ee
A plot of this orbital velocity
for the exact $U(x) $ and the Newtonian approximation 
is given in figure \ref{fig:V}, for
$\omega = 0.1, \, g_0 = 1, \d = 0.1$ thus $U_0 = - 0.005$.
Note that these simple relations hold only outside 
of the mass distribution, and will be modified by the
presence of mass e.g. in the halos of galaxies, 
and by the onset of the harmonic long-distance decay
\eq{harmonic-decay-v}. 
The result is similar to that of an Schwarzschild-de Sitter geometry
with $\Lambda_{\rm eff}$ given by \eq{Lambda-eff}, 
combined with a harmonic
screening at $r \approx L_\omega$.
It remains to be seen if this indeed allows to explain the 
observed galactic rotation curves; however qualitatively, the
above behavior certainly goes in the right direction.
What is particularly striking is that it naturally predicts a 
slightly increasing rotation curve, 
which is indeed often observed. This may be masked by other
effects, in particular the non-trivial matter distribution.
In fact, the possibility that the galactic rotation curves might be 
explained in terms of a cosmological constant 
has been proposed in the literature \cite{Whitehouse:1999rs}.
It was argued to be feasible, provided $\Lambda_{\rm eff}$ can depend 
on the individual galaxy 
(the value $\Lambda_{\rm eff} \approx - 5 \times 10^{-55} cm^{-2}$ 
was given as a typical scale). 
Of course  this does not make sense in conventional GR,
and it is inconsistent with cosmological constraints in the
$\Lambda$CDM model.
However the latter are irrelevant here as discussed below, and 
$G =  \frac{g_0^2\omega^4}{\Lambda_1^4}$ is dynamical here
and may indeed depend on the 
individual galaxy. It would be extremely
interesting to study this in more detail.

\paragraph{Some estimates.}

We certainly want to preserve $g_{00} <0$,
which implies that
\be
1 > 2U_{0} = g_0^2 \omega^2 = G \frac{\Lambda_1^4}{2\omega^2} 
=  \frac{\Lambda_1^4}{2\omega^2\Lambda_{\rm planck}^2} .
\ee
This allows to express the vacuum energy as
\be
\Lambda_1^2 = \sqrt{2U_0}\, \omega \Lambda_{\rm planck} 
 =  \sqrt{2U_0}\, \frac {\pi}{L_\omega L_{\rm planck}}
\ee
To get some idea of the scales we consider our galaxy. 
Assuming $L_\omega \approx L_{\rm galaxy} \approx 10^{20} m$ 
(so that the oscillations occur roughly at the size of 
our galaxy), this gives
\be
\Lambda_1 \approx  U_0^{1/4}\, \sqrt{\frac 1{10^{20}m  10^{-35}m}}
\approx  U_0^{1/4}\, 10^7 m^{-1} \approx  U_0^{1/4}\, 10 [eV]. 
\ee
To be specific assume that the 
potential energy at the center or the galaxy 
roughly coincides with its value given by Newtonian gravity,
\be
U_0 \approx M G/L_{\rm galaxy} \approx \frac{10^{15}m}{10^{20}m}
\approx 10^{-5}
\ee 
where $M \approx 10^{41} kg$.
Then we get a bound on $\Lambda_1$
of the order of electronvolt.
We will see in section \ref{sec:cosmology}
that the cosmology in this model 
is not very sensitive to the actual value of 
$\Lambda_1$. Moreover, $\Lambda_4$
could be {\em much} smaller than the Planck scale
in this model as discussed below. Thus the
fine-tuning is considerably milder
than in the standard $\Lambda$CDM model
where  $\Lambda_1 \approx 2 \times 10^{-3} eV$.

\section{General matter distribution}
\label{sec:general-rho}

In this section, we generalize the above considerations to the 
case of an arbitrary but still somewhat localized mass distribution.
The approach is a bit different from the one on the previous section, 
which also provides a confirmation of the above results.

\begin{figure}
  \vspace{-0.2cm}
  \hspace*{1cm}
 \includegraphics[scale=0.55]{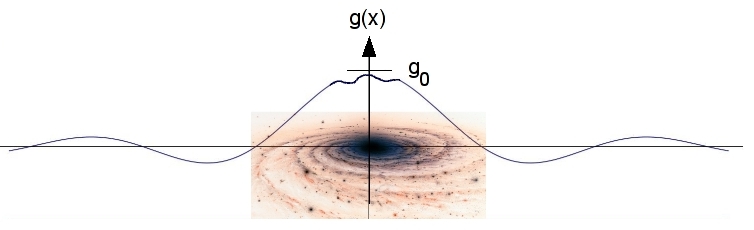}
\caption{sketch of embedding function $g(x)$ with short-scale perturbations and 
long-distance oscillations}
\label{fig:galaxy}
\end{figure}

Consider a large harmonic ``gravity bag'' solution 
\eq{phi-0} with size $L_\omega$ 
of the order of a galaxy, say, which satisfies
\bea
\phi^i_0(x,t) &=& g_0(x) e^{i\omega t} ,\nn\\
\Delta g_0(x) &=&  -\omega^2 g_0(x) .
\label{vacuum-bag-2}
\eea
Now add (non-relativistic)
matter with $\rho(x) =\tilde T_{00} \geq 0$ and  $\tilde T_{ij} \approx 0$
into this bag; 
indeed matter will tend to accumulate in the bag due to its 
attractive gravitational potential \eq{Delta-U-vac}.
If the localized matter density $\rho(x)$ (consisting of stars, etc.) 
is not too large,
this will lead to a deformation 
\bea
\phi^i(x,t) &=& g(x) e^{i\omega t} , \nn\\
g(x) &=& g_0(x) + \d g(x)
\label{harmonic-ansatz-general}
\eea
where $|\d g| \ll g_0(x) \approx g_0(0)$ is varying on short scales
according to $\rho(x)$, while $g_0(x)$ is slowly 
varying at the scale $L_\omega$. 
Notice that this is precisely the splitting in \eq{g-r-phase},
and the condition $|\d g| \ll g_0(x)$ 
corresponds to \eq{outside-condition}.
Thus $g_0(x)$ is the 
embedding due to the average, overall mass in the galaxy resp. 
the largest cosmic structures, and $\d g(x)$
due to small local perturbations such as stars, as indicated
in figure \ref{fig:galaxy}.
This dominance of large-scale structures is vindicated 
by the fact that 
the typical velocities of large-scale structures
in cosmology are larger than for small-scale structures,
cf. \eq{v-SdS}.

\subsection{Metric deformation and Newtonian limit}

In this situation, it makes sense to
linearize in $\d g(x)$; this is the crucial step. 
The corresponding metric is static and has the form 
\eq{basic-metric-static}
\be
d s^2 = -(1- \omega^2 \, g^2) \, dt^2 + 
(\d_{ij} + \partial_i g \partial_j g) dx^i dx^j .
\ee
We want to derive the Poisson equation for the 
corresponding gravitational potential
$U(x) = -\frac 12 \omega^2 g^2$. Thus consider 
\be
(\Delta + \omega^2) U = -  \omega^2 g(x) (\Delta+\omega^2) g(x) 
- \omega^2 \d^{ij} \partial_i g \partial_j g .
\ee
Now recall the equation of motion \eq{eom-general-linear-rho} 
\be
(\Delta+\omega^2) g =  -8\pi\omega^2\frac{\rho}{\Lambda_1^4} g(x) .
\label{eom-embedd-1}
\ee
Using \eq{vacuum-bag-2} this gives for the fluctuations 
\be
(\Delta + \omega^2)\d g =  - 8\pi\omega^2\frac{\rho}{\Lambda_1^4} (g_0 + \d g)
\approx  - 8\pi\omega^2\frac{\rho}{\Lambda_1^4} \, g_0 .
\label{eom-embedd-2}
\ee
Now we use the assumption that $\rho$ is some small mass 
distribution inside a large 
background bag $g_0$ which is slowly varying. This
means that 
\bea
\partial_i g_0 &=& O(r \Delta g_0) = O(r \omega^2 g_0), \nn\\
\partial_i \d g &=& O(\tilde r \Delta \d g)
= O(\tilde r\omega^2 \frac{\rho}{\Lambda_1^4} \, g_0)
\eea
where $r \ll L_\omega$ denotes the distance from the
``center'' of the bag, and $\tilde r$ is the distance from $\rho$.
Putting this together and using $\omega^2 r^2 \ll 1$, 
we obtain 
\bea
(\Delta + \omega^2) U &=& 8\pi\omega^4 g(x)^2 \frac{\rho(x)}{\Lambda_1^4}
- \omega^2\d^{ij} \partial_i g \partial_j g \nn\\
&=&  8\pi\omega^4 g_0^2 \frac{\rho(x)}{\Lambda_1^4}
\,+ O\Big(r^2 \omega^2 \omega^4 g_0^2 
(1+8\pi\frac{\rho}{\Lambda_1^4})\Big)\, + O(\d g^2) 
\eea
This shows that the terms with first derivatives 
can be neglected
assuming $r^2 \omega^2 \ll 1$, and the term $O(\d g^2)$ 
can be neglected for small $\rho$. 
Identifying the Newton constant as
\be
G = 2g_0^2\,\frac{\omega^4}{\Lambda_1^4} ,
\label{G-def-0}
\ee
we  obtain the Poisson equation
\be
\Delta U \approx  4\pi G\, (\rho(x) + \frac{\Lambda_1^4}{8\pi})
\ee
with an effective vacuum energy $\Lambda_1^4$ which could be 
interpreted as apparent negative cosmological constant 
$\Lambda_{\rm eff}$.
This confirms the results  \eq{Newton-constant}, \eq{Lambda-eff}
of the previous section.
If the matter density is much larger than the
vacuum energy,
\be
\frac{\rho}{\Lambda_1^4} \gg 1 
\ee
this gives indeed the usual Poisson equation of Newtonian gravity,
\bea
\Delta U &=&  4\pi G\, \rho .
\eea 
As a check, consider the case of a small localized mass $M$ at the origin.
The fluctuations $\d g \ll g_0$ can be determined explicitly
using \eq{eom-embedd-2},  which implies
$(\Delta + \omega^2)\d g \approx  -4\pi \frac{G \rho}{g_0 \omega^2}$. 
Outside of the mass distribution this gives
\bea
\d g &=& g_0\a \, \frac{\sin(\omega r+\tilde\d)}{\omega r}  , \nn\\
g &=& g_0 + \d g = g_0\Big(\frac{\sin(\omega r)}{\omega r} 
+ \a \,\frac{\sin(\omega r+\tilde\d)}{\omega r}\Big)
\,\approx\,  g_0\frac{\sin(\omega r+\d)}{\omega r}, \nn\\
\frac{G M}{\omega g_0^2} &=&  \a \sin(\tilde\d) .
\eea
This results in a metric which is essentially the same as
\eq{g00-simple}, \eq{grr-simple},
 replacing $\d = \a\tilde\d$ for small $\a$.
However, a more complete treatment including corrections 
due to $\theta^{\mu\nu}(x)$ and more general 
``gravity bags'' is required before
 a detailed comparison with the solar
system constraints can be performed.

We conclude that localized matter $\rho$ inside the gravity bag $\phi_0$
is subject to (emergent) Newtonian gravity, with a dynamically
determined  gravitational ``constant'' $G$ given by \eq{G-def-0}.
For example, two stars or planets would lead to 
local perturbations
\be
g(x) = g_0 + \d g_1(x) + \d g_2(x)
\ee
where $\d g_i$ are perturbations due to 
object $\rho_i$. They both see the same $g_0$ and $\omega$, 
thus the same gravitational constant $G$, and 
Newtonian gravity is recovered 
on scales shorter than $L_\omega$.
However at very long scales $L_\omega$ and 
near the border of the galaxy, the effective gravitational
constant might differ; this would require more detailed modeling.

Recall that
this mechanism is necessarily non-linear and therefore
somewhat non-trivial, but nevertheless it is 
quite robust: 
the essential ingredient is the gravity bag 
$g_0(x) e^{i\omega t}$
which must be slowly rotating and have large amplitude.
As discussed before, it seems unavoidable that 
such a bag forms around large matter clusters in the static case. 
This leads to an effective gravity which is at least close to
what we observe, and might provide an explanation for the 
rotational curves in typical galaxies without 
resorting to ``dark matter''. 
An obvious question is how the parameters
$\omega$ and $g_0$, and thereby $G$ are 
determined; we will give an argument below why 
$G$ should be at least 
approximately the same in different galaxies.

\subsection{Ricci tensor and relation with general relativity}

In order to compute the Ricci tensor in the above
situation, we assume for simplicity that the mass distribution
$\rho$ is in the center of the background gravity bag, so that 
\bea
\partial_i \phi_0 &\approx& \partial_i g_0 = 0, \nn\\
\partial_i \partial_j g_0 &\approx&  
 -\frac 13 \delta_{ij} \omega^2  g_0
\label{background-coords-2}
\eea
and $g_0(x) \approx g_0 -\frac 16 \omega^2 g_0 r^2$ 
for small distances. The space-like components of the metric
$h_{ij} =  \partial_i g \partial_j g$ can then be written as
\bea
h_{ij} &=&  \partial_i g_0 \partial_j g_0   
+ 2\partial_i g_0 \partial_j \d g  + O(\d g^2) \nn\\
 &=&  \frac 19 g_0^2 \omega^4  x_i x_j
- \frac 23 g_0 \omega^2 x_i \partial_j \d g  + O(\d g^2) .
\eea
The Ricci tensor can be computed
using the on-shell relation \eq{h-gauge}
\be
\partial^\l h_{\l\nu} = \frac 12 \partial_\nu h + \Box\phi (\partial_\nu\phi) 
= \frac 12 \partial_\nu h - 8\pi\omega^2 \frac {\rho}{\Lambda_1^4} \phi (\partial_\nu\phi) 
\ee
hence
\bea
\partial_\mu\partial^\l h_{\l\nu} &=& \frac 12\partial_\mu\partial_\nu h 
+ \Box \phi (\partial_\mu\partial_\nu\phi) 
+ \partial_\mu(\Box \phi) \partial_\nu\phi 
\eea
so that
\bea
R_{\mu\nu} &=& \partial_\a \Gamma^\a_{\mu\nu} - \partial_\mu
\Gamma^\a_{\a\nu} 
= \frac 12 \(-\Box h_{\mu\nu} - \partial_\mu\partial_\nu h
+ \partial_\mu \partial^\l h_{\l\nu} + \partial_\nu \partial^\l
h_{\l\mu}\) 
\nn\\
&=& -(\partial_\mu\partial^\l\phi)(\partial_\nu\partial_\l\phi)
 + (\Box\phi) (\partial_\mu\partial_\nu\phi) \nn\\
R &=& -(\partial^\mu\partial^\l\phi)(\partial_\mu\partial_\l\phi)
+ (\Box\phi)^2 \nn\\
&=&  \omega^4 g^2 - 2\omega^2 \partial^\l \phi \partial_\l \phi
 - \partial^i\partial^j g \partial_i\partial_j g
- (8\pi\omega^2 \frac {\rho}{\Lambda_1^4})^2 g^2 
\eea
(using $\partial_0\phi \partial_0\phi = \omega^2 \phi^2 =
-\phi \partial_0^2\phi$ and \eq{background-coords-2}).
This is hard to evaluate in general. 
We can write it as sum of contributions 
$R_{\mu\nu} = R_{\mu\nu}^{(0)} +  R_{\mu\nu}^{(1)}+ R_{\mu\nu}^{(2)}$
where
\bea
R_{\mu\nu}^{(0)} &=&  -(\partial_\mu\partial^\l\phi_0)
(\partial_\nu\partial_\l\phi_0) \nn\\
R_{\mu\nu}^{(1)} &=&  -(\partial_\mu\partial^\l\phi_0)(\partial_\nu\partial_\l\d\phi)
-(\partial_\nu\partial^\l\phi_0)(\partial_\mu\partial_\l\d\phi)
+ (\Box\d\phi) (\partial_\mu\partial_\nu\phi_0) \nn\\
R_{\mu\nu}^{(2)} &=&  -(\partial_\mu\partial^\l\d\phi)
(\partial_\nu\partial_\l\d\phi)
 + (\Box\d\phi) (\partial_\mu\partial_\nu\d\phi) 
\eea
since $\Box\phi_0=0$.
$R_{\mu\nu}^{(0)}$ is the ``vacuum'' contribution due to the 
background bag,
$R_{\mu\nu}^{(1)}$ is the desired contribution 
due to matter linear in $\rho$,
and $R_{\mu\nu}^{(2)} = O(\rho^2)$ is expected to be negligible
for small $\rho$.
Consider first the ``vacuum contribution'' $R_{\mu\nu}^{(0)}$
which applies whenever $\rho=0$:

\paragraph{Harmonic bag contribution.}

The contribution of the harmonic bag $\phi_0$
is given by 
\be
R_{\mu\nu}^{(0)} = -(\partial_\mu\partial^\l\phi_0) (\partial_\nu\partial_\l\phi_0) , 
\qquad R = -(\partial^\mu\partial^\l\phi_0) (\partial_\mu\partial_\l\phi_0).
\ee
This can be written 
using $\partial_0\phi \partial_0\phi = \omega^2 \phi^2 =
-\phi \partial_0^2\phi$ and \eq{background-coords-2}
as
\bea
R_{00} &=& -\omega^2 \partial^\l\phi_0 \partial_\l\phi_0 \approx 
- \omega^4 g_0^2 = -\frac 12 G \Lambda_1^4, \nn\\ 
R_{ij} &=& - (\partial_i\partial_k\phi_0) (\partial_j\partial_k\phi_0)  
 \approx - \frac 19 \d_{ij} \omega^4 g_0^2 
=  - \frac 1{18} \d_{ij}  G \Lambda_1^4\nn\\
R &\approx&  \frac 23 \omega^4 g_0^2 = \frac 13 G \Lambda_1^4 .
\eea
Therefore $R_{\mu\nu}^{(0)} = O(G\Lambda_1^4)$ as expected, and
\bea
\cG_{00} &=& R_{00} - \frac 12 \eta_{00} R 
= -\frac 13  G \Lambda_1^4,   \nn\\
\cG_{ij} &=& R_{ij} - \frac 12 \d_{ij} R 
= - \frac 2{9} \d_{ij}  G \Lambda_1^4 .
\eea
In the presence of some
local matter distributions $\rho(x)$, this background will be  
dominated by the gravitational field due to $\rho(x)$.
This is contained in $R_{\mu\nu}^{(1)}$
which we compute next.

\paragraph{Linear matter contribution.}

Now consider
$R_{\mu\nu}^{(1)}$, which is the desired contribution linear in
$\rho$. 
To proceed we consider the time-and space-like components separately:
\bea
R_{00} &=& -2\omega^2 \partial^\l\phi_0 \partial_\l\d\phi
+  8\pi\omega^4 g_0^2 \frac {\rho}{\Lambda_1^4} \nn\\
&=& -2 \omega^4 g_0\d g  +  8\pi\omega^4 g_0^2 \frac {\rho}{\Lambda_1^4}
\approx  4\pi G \rho  + 2 U(x) \omega^2  \nn\\
&\approx&  4\pi G \rho 
\label{R00-rho}
\eea
where $U(x)$ is the gravitational potential due to $\rho$
as determined earlier; that term is negligible since
$4\pi G \rho \approx \Delta U(x) \gg U(x) \omega^2 $
by assumption. This is in agreement with GR as discussed below.
 $R_{0i} = 0$ follows again from $\partial_0 \phi \partial_i\phi = 0$,
and 
\bea
R_{ij} &=& -(\partial_\mu\partial^\l\phi_0)(\partial_\nu\partial_\l\d\phi)
-(\partial_j\partial^\l\phi)(\partial_i\partial_\l\d\phi)
-  8\pi\omega^2 \frac {\rho}{\Lambda_1^4} g_0 \partial_i\partial_j g_0\nn\\
&\approx&  -(\partial_i\partial_k g_0)(\partial_j\partial_k\d g)
-(\partial_j\partial_k g_0)(\partial_i\partial_k\d g)
-  8\pi\omega^2 \frac {\rho}{\Lambda_1^4} g_0 \partial_i\partial_j g_0\nn\\
&\approx& \frac 23 \omega^2  g_0 \partial_j\partial_i\d g
+ \frac 13 4\pi G\rho \d_{ij} 
\eea
using \eq{background-coords-2}. This gives
\bea
R &=& -   4\pi G \rho +\frac 23 \omega^2  g_0 \Delta\d g
+ 4\pi G \rho 
= - \frac 23  4\pi G \rho 
\label{R-matter}
\eea
and
\bea
\cG_{00} &=& R_{00} - \frac 12 \eta_{00} R 
= \frac 23  8\pi G \rho ,   \nn\\
\cG_{ij} &=& R_{ij} - \frac 12 \d_{ij} R 
\approx \frac 23 \omega^2  g_0 \partial_j\partial_i\d g
+ \frac 23 4\pi G\rho \d_{ij} .
\label{G-T-equal}
\eea
In particular, in regions where $\rho(x)=0$ 
we obtain indeed
\be
\cG_{00} = R_{00} = G T_{00} = 0, \quad R = 0
\label{G00-vac}
\ee
as in GR, up to corrections due to the vacuum energy. 
Recall that in GR, 
the Einstein equations for $T^{00} =
\rho,\,\,T^{ij} = 0$ imply $\cG_{00} =8\pi G \rho$, and
furthermore 
$R_{ij} = 4\pi g_{ij} G \rho,
\,\, R_{00}  = 4\pi G\rho$
and $8\pi G\rho = R$. Thus \eq{R00-rho} and
\eq{G00-vac} agree with GR, while the space-like
(pressure) components of $R_{ij}$ and $\cG_{ij}$
are different here from GR.
In particular, for $\rho=0$ the essential difference to GR is 
 an anisotropy of $R_{ij}$.
Indeed we should  not expect complete agreement with GR
since there are no  harmonic embeddings of non-trivial
Ricci-flat 4-manifolds \cite{nielsen}.

The non-vanishing components of $\cG^{ij} = O(G\rho)$ 
would endanger the Newtonian limit if indeed $\Lambda_4 = \Lambda_{\rm planck}$.
On the other hand, this does not pose any problem if 
$\Lambda_4 \ll \Lambda_{\rm planck}$,
so that the  contributions of $\Lambda_4^2 \cG^{ij}$ are negligible.
This scenario is quite possible and in fact very appealing
as discussed below, greatly
reducing the required fine-tuning for $\Lambda_1$.

\subsection{Quantization, scales and stability}
\label{sec:stability}

Perhaps the most remarkable result 
is that gravity arises in the
harmonic branch of the matrix model, without even using the 
induced gravitational action i.e. the Einstein-Hilbert term.
This means that the scale $\Lambda_4$ in front of the induced 
Einstein-Hilbert term \eq{basic-action} could 
be much smaller than the Planck scale,
\be
\Lambda_4 \ll \Lambda_{\rm planck}
\label{planck-L4}
\ee
because the Newton constant $G$ is determined
dynamically  through \eq{Newton-constant}.
In principle, $\Lambda_4$ might be as low as $O(TeV)$ \footnote{
A similar possibility has also been considered 
in \cite{ArkaniHamed:1998rs}
however in a rather different context of higher-dimensional
GR with branes.}.
In fact this {\em should} be so in the present context, since 
otherwise the non-vanishing components $\cG_{ij}$ found above
would enter equation \eq{eom-general} 
and might spoil the results of the previous sections
(unless they vanish in a more sophisticated solution).
This is of course  a very attractive scenario,
because then the difference of scales between 
$\Lambda_1$ and $\Lambda_4$ would be greatly reduced,
largely resolving the cosmological constant problem.

Another extremely interesting feature of the harmonic branch 
of the model is that
the quantization of gravity seems to be straight-forward
and well-behaved. 
Indeed $\Lambda_4$ is essentially the scale of 
$N=4$ SUSY breaking, and the model can be viewed as non-commutative
$N=4$ SYM on $\R^4_\theta$ which is expected to 
provide a well-defined quantum theory.
The degrees of freedom in $\theta^{\mu\nu}(x)$
can be interpreted as would-be $U(1)$ non-commutative
gauge fields on $\R^4_\theta$, which are also well-behaved
in the IKKT model. The excitations of the scalar
fields $\phi_i$ are essentially harmonic excitations
due to the brane tension/vacuum energy
\be
 S_{\rm vac} =  -2\Lambda_1^4\int d^4 x \sqrt{|g|}
 \approx -\Lambda_1^4\int d^4 x 
(2+ \eta^{\mu\nu}\partial_{\mu}\phi^i \partial_\nu\phi^j\d_{ij})
\ee 
with positive energy
\be
E_{\rm vac} \approx \Lambda_1^4\int d^3 x\,
 (2+\partial_0\phi\partial_0\phi + \partial_i\phi\partial_i
 \phi\d^{ij}) .
\label{E-vac-2}
 \ee
This is similar to the quantization of a Klein-Gordon field.
Note in particular that the energy of the gravity bags
is proportional to $g_0^2 \omega^2$, hence its scale is 
more-or-less determined dynamically by the 
initial energy distribution. 
In particular, it is plausible that the total energy 
of the universe consisting of 
brane fluctuations and the total mass is positive.
This suggests that also after structure formation, the 
(vacuum) energy of the gravity bags associated to a 
galaxy with mass $M$ should be no less than
its gravitational binding energy $M U$.
Then increasing $g_0$ would increase the total energy, which is
important for stability reasons.

In contrast, many of the difficulties associated with the 
quantization of GR are expected to be 
recovered in the Einstein branch of the model.

It remains to be seen whether the emergent gravity 
is compatible with the
precise solar system constraints, which requires  a
more complete analysis.
Finally it should be pointed out that the
considerations of this paper should be extended to 
cover the case of small extra dimensions, e.g. 
in the form of fuzzy sphere(s). The basic
results of this paper are expected to apply also 
in that case.

\subsection{Cosmological context and perturbations}
\label{sec:cosmology}

We briefly explain how the 
above solutions fit into a consistent cosmology.
Assuming that the vacuum energy $\Lambda_1^4$ dominates the 
energy density due to matter,
cosmological solutions of emergent NC gravity were
obtained in \cite{Klammer:2009ku} as harmonically embedded
branes $\cM^4 \subset \R^{10}$ 
\be
\vec x(t,\chi,\theta,\varphi) = \(\begin{array}{c}
\cR(t) \(\begin{array}{l} \sinh(\chi)\sin\theta\cos\varphi \\
     \sinh(\chi)\sin\theta\sin\varphi  \\
     \sinh(\chi)\cos\theta \\
     \cosh(\chi) \end{array}\)\\
0 \\  x_{c}(t)\end{array}\) \in \R^{10} \nn
\ee
where
\be
\cR(t) = a(t)\,\(\begin{array}{l} \cos\psi(t) \\
\sin\psi(t) \end{array}\) .
\label{R-rotation}
\ee
Here $\eta_{ab} =\diag(+,...+,-,-,+,+)$ for $k=-1$.
This embedding is harmonic, $\Delta_g \vec x =0$, and 
has a FRW geometry
\be
ds^2 = -dt^2 + a(t)^2 d\Sigma^2,  \qquad
d\Sigma^2 = d\chi^2 + \sinh^2(\chi)d\Omega^2 
\label{FRW}
\ee
corresponding to spatial curvature $k=-1$.
The physics of the early universe in this model
is quite different from standard cosmology and 
requires a more detailed analysis\footnote{
In particular the consistency with the CMB data 
cannot be reliably addressed at this point, 
see however \cite{BenoitLevy:2009yf} for the simplified case of 
an exact Milne universe. The above refined solutions 
yield a big bounce and 
an early phase with power-law acceleration \cite{Klammer:2009ku},
but a proper treatment of matter is still missing.};
however,  it is a solid prediction that the solution 
approaches $a(t) \to t$ for late times, i.e. a 
Milne universe. Remarkably, this geometry is in 
good agreement with the basic observational 
data including the tye Ia supernovae data \cite{BenoitLevy:2009yf}, 
which are usually interpreted in terms of an accelerating
universe.

\paragraph{Milne universe.}

The Milne geometry is expected to hold in this model 
as long as the vacuum energy $\Lambda_1^4$ dominates the 
energy density due to matter. In particular,
there is no upper bound
on $\Lambda_1$ from cosmology. 
These statements are easy to understand, recalling that
the Milne universe is nothing but (a sector of) 
flat Minkowski space $\R^4 \subset \R^{10}$.
Indeed consider the flat metric 
$d s^2 = -d\tau^2 + dr^2 + r^2 d\Omega^2$ on $\R^4$. 
In terms of the new variables 
\be
\tau =  t \cosh(\chi), \,\, r =  t \sinh(\chi),
\ee
this metric takes the form of a FRW metric with 
$a(t) = t$ and $k=-1$,
\bea
d s^2_g &=& -d\tau^2 + dr^2 + r^2 d\Omega^2
  = \frac {a^2}{\tilde t^2}\(-d t^2 +  t^2(d\chi^2 +
\sinh^2(\chi)d\Omega^2)\) \nn\\
&=& -  d t^2 +  a(t)^2 (d\chi^2 +\sinh^2(\chi)d\Omega^2) .
\eea
Clearly,  flat Minkowski space is a solution 
of a brane with tension resp. vacuum energy $\sim \Lambda_1^4$,
and in fact it is stabilized by a large $\Lambda_1$.

We claim that matter such as stars and galaxies lead to local
perturbations resp. fluctuations 
of this cosmological solution, and the previous results 
including Newtonian gravity and long-distance deviations
go through with very minor adaptation.
This can be seen easily by using
the previous localized solutions on $\R^4$ and rewriting them 
in terms of the Milne variables. For example, the spherically
symmetric solutions centered at the origin $r=\chi=0$ become  
\bea
\phi_0(x) &=& g(r) e^{i\omega \tau} 
= g_0 \frac{\sin(\omega r)}{\omega r} e^{i\omega \tau}  \nn\\
&=& g_0 \frac{\sin(\omega t \sinh(\chi) )}{\omega t \sinh(\chi) } 
e^{i\omega \cosh(\chi) t}  \nn\\
&\approx& g_0 \frac{\sin(\omega a(t) \chi )}{\omega a(t) \chi} 
e^{i\omega t}  
\eea
using $\chi \ll 1$ in the neighborhood of the mass. Thus in
comoving Milne coordinates, this looks like a 
spherically symmetric solution with red-shifted wavelength 
$a(t) \omega = t \omega $. Note that the effective gravitational
constant $G= \frac{2 g_0^2\omega^4}{\Lambda_1^4}$ 
is independent of $t$, which is reassuring\footnote{$G$ might 
change in the very early universe due to deviations from the Milne 
metric.}.
A particularly  interesting aspect is the slow ``extrinsic''
 rotation \eq{R-rotation}
in the embedding $\cM \subset \R^{D}$ for small 
$t$, which turns out to be \cite{Klammer:2009ku}
\be
\dot\psi \sim \frac 1{a^5} .
\ee
This might be very important in the context of fluctuation 
spectra in the early universe.
Clearly  a more complete perturbation analysis is required, 
in particular in order to address the issues of 
fluctuation spectra in the context of the cosmic microwave
background.

We can thus summarize the physical aspects of the model 
as follows, leaving aside its theoretical appeal e.g. with respect to 
quantization. The most attractive feature of the model is 
that it naturally predicts a (nearly-) flat universe, resulting in
luminosity curves e.g. for type Ia supernovae which are close 
to the observed ones
(usually interpreted in terms of cosmic acceleration)
without any fine-tuning.
On the other hand, reconciling it with observed gravity 
on scales less than or equal to galactic scales does 
impose an upper bound on the vacuum energy
of order eV. This in turn might offer a mechanism for 
(partially?) explaining
the galactic rotation curves, without requiring large 
amounts of dark matter.

\paragraph{Gravitational (non--)constant $G$.}

Up to now, the parameters $\omega, g_0$ 
and therefore $G$ were undetermined. 
However these are dynamical quantities, in particular
they depend on the initial conditions. 
Thus we have to face the question why in particular $G$ should be 
the same in different parts of the universe, in particular
in different galaxies. Indeed one should expect  in this model that 
$G$ depends somewhat on the individual galaxies,
however the variation should certainly not be too large 
in our universe.

While we cannot offer a completely satisfactory answer yet, some 
insight can be obtained from the above cosmological solutions. 
We have seen that deformations $g(x)$ of the brane 
due to e.g. galaxies can be considered as 
perturbations propagating in an approximately 
flat Milne resp. Minkowski background, with
$\omega$ and $g_0$ remaining essentially constant in time.
In reality, we know that galaxies are typically 
parts of larger structures such as (super)-clusters and filaments.
This large-scale structure is expected to provide the 
dominant contribution to $g(x)$ and therefore to 
$G$, which is therefore related over cosmological scales. 
Moreover, it is very plausible that the dynamics of structure
formation in the universe leads to similar scales for $g_0(x)$
of these dominant cosmic structures, given the 
high degree of homogeneity of the initial conditions 
seen in the CMB background. Furthermore, the rotation
$\dot \psi$ of the above cosmological solution \eq{R-rotation}
might provide a natural seed for the required 
rotation of the large-scale gravity bags.
On the other hand, there could be some other
more rigid mechanism for stabilizing $G$ which 
we did not identify here.

\section{Discussion}

We obtained in this paper some remarkable 
results on the long-distance properties 
of emergent gravity in Yang-Mills matrix models,
 notably the IKKT model. 
Space-time is modeled by a 3+1-dimensional
noncommutative brane solution, which acquires an
``emergent'' metric. Its dynamics
is governed by the brane tension 
as well as additional gravitational terms including the Einstein-
Hilbert term in the (quantum) effective action.
There are two types of gravitational
solution: an ``Einstein branch''
which is very similar to general relativity, and a 
``harmonic branch'' where the branes are governed by a brane
tension. We focus on the harmonic branch in this paper, 
and study the 
deformation of the space-time brane and its effective metric due to 
static localized mass distributions, having in mind e.g. galaxies.

Due to the brane tension, the basic excitations of the 
space-time brane are
essentially harmonic waves. 
Large matter clusters such as galaxies are embedded 
in such ``gravity bags'', which are rotating standing waves
of the embedding with long wavelength $L_\omega$.
Standard Newtonian gravity is recovered 
inside these ``gravity bags'' due to local matter such as stars. 
Moreover, there is an effective gravitational constant 
$\Lambda_{\rm eff}$ inside the bags, which leads 
to a significant enhancement of orbital velocities
at large distances,  quite reminiscent of the 
observed galactic rotation curves. 
At very large distances $\geq L_\omega$, the harmonic embedding
leads to a 
screening $\sim \frac 1{r^2}$ of the gravitational potential, 
reconciling relatively large vacuum energies (of order
$eV$, say) with a consistent cosmology similar to a Milne 
universe. The latter is known to be in remarkably good
agreement with the basic cosmological constraints,
leaving aside the CMB fluctuations which 
require a detailed understanding of the early universe in this model.

These results have important physical implications.
An obvious conjecture is that the observed enhancement of the
galactic rotation velocities compared with the Newtonian law
is primarily due to the above results, 
i.e. an effective vacuum energy
$\Lambda_{\rm eff}$
inside the gravity bags. This gets additional support by the 
result that the gravitational constant $G$ 
and therefore $\Lambda_{\rm eff}$
is not universal but determined
dynamically, and may therefore differ somewhat from galaxy to galaxy.
This does not imply that there is no dark matter at all, but
the model clearly requires far less dark matter in the 
galactic halos than what is invoked in the $\Lambda$CDM model.

It is certainly remarkable how close this simple and rigid model 
comes to observation.
In particular, large-scale cosmological observations
seem to be reproduced much more naturally 
in the harmonic branch than in GR, 
without particular fine-tuning of the vacuum energy.
The basic solar system observations can be reproduced 
to a good approximation, which seems to require an 
upper bound on $\Lambda_{\rm eff}$ on the 
order of $eV$.
However, the precision tests provide a challenge,
and it remains to be seen whether they can be met 
in the harmonic branch of solutions under consideration.
The only obvious modifications of the matrix model would be 
quadratic or cubic (soft) terms,
 which do not spoil the good UV behavior of its quantization. 
The cubic terms lead to compactified extra dimensions
as fuzzy spheres, which is very natural and desirable
for particle physics \cite{Aschieri:2006uw}.
The quadratic terms might help to 
stabilizes the ``scale'' of space-time
resp. the effective vacuum energy. Indeed
observational constraints require
that the vacuum energy is quite small (perhaps $O(eV)$).
This may be due to a cancellation between the bare action 
and the induced quantum-mechanical vacuum energy, but
quadratic terms in the matrix model might also
play an important role here.

Another remarkable feature of the model is that gravity
is naturally weak. Indeed the scale of gravity is found to be 
$G \sim \frac{g_0^2}{L_\omega^4\Lambda_1^4}$, where
$L_\omega$ is a cosmological length scale, and 
$\Lambda_1$ the vacuum energy. Moreover $G$ can
differ to some extent in different parts of the universe,
and depends on the surrounding mass distribution which determines 
a ``gravity bag''.

In any case, an essential result of this paper 
is that Newtonian gravity arises in the matrix model
simply due to the brane tension
of a harmonically embedded space-time brane;
the Einstein-Hilbert action is not required.
This suggest the following very appealing scenario: 
the coefficient $\Lambda_4^2$ of the induced 
Einstein-Hilbert term could be much {\em smaller} 
than the Planck scale. 
This would also greatly alleviate the 
fine-tuning issue for the vacuum energy, and 
the precise form of the (induced) gravitational action 
is not essential. 
This basic mechanism 
might apply in a more general framework than that of
Yang-Mills matrix models. However, the matrix model
offers a clear concept of quantization,
in particular the IKKT model.  It contains
not only  (emergent) gravity but all ingredients required
for a theory of fundamental interactions,
in particular nonabelian gauge fields and fermions. 
This would provide a very interesting 
and accessible quantum theory of
gravity, which has the potential to resolve some fundamental
problems in this context. 

There are many obvious shortcomings in this paper
which require more detailed work. 
The present understanding is still at a rather crude level, 
and more detailed work is needed before claiming 
to seriously challenge the $\L$CDM model.
Probably the main problem are the space-like components
of $\cG^{\mu\nu}$ resp. $g_{rr}$, which do not quite 
agree with GR even in vacuum.
It remains to be seen if the solutions presented here 
can meet the solar system precision tests, or if a more
complete solution taking into account notably $\theta^{\mu\nu}(x)$
will lead to the suitable corrections.
Furthermore, 
the short-distance properties of a point-mass solution
i.e. black holes should be studied in the matrix model 
framework. Here higher-order corrections 
due to $\theta^{\mu\nu}$
are expected to play an important role.
Finally, the Einstein branch of the IKKT model 
should provide a quantization of
 more conventional
general relativity coupled to matter.

\paragraph{Acknowledgments}

The author wishes to thank in particular N. Arkani-Hamed 
for very useful discussions and hospitality at the 
IAS Princeton, as well as
R. Brandenberger, G. Dvali, 
H. Grosse, D. Klammer, H. Rumpf and I. Sachs for useful 
discussions, and 
the CERN theory division for hospitality. 
This work was supported 
in part by the FWF project P20017 
and in part by the FWF project P21610.

\end{document}